\documentclass[pra,twocolumn,aps,superscriptaddress,longbibliography]{revtex4-1}

\usepackage[normalem]{ulem}
\usepackage{amsmath}
\usepackage{amssymb}
\usepackage{lineno}
\usepackage{graphicx}
\usepackage[usenames,dvipsnames]{color}
\usepackage{braket}
\usepackage{bm}
\usepackage{mathtools}
\usepackage{times}
\usepackage{hyperref}
\usepackage{pdfpages}
\usepackage{float}
\usepackage{lipsum}
\usepackage[dvipsnames]{xcolor}
\hypersetup{
  colorlinks=true,
  citecolor=blue,
  linkcolor=blue,
  urlcolor=blue}
\usepackage{tikz}
\usetikzlibrary{decorations.markings}
\usetikzlibrary{decorations.pathmorphing,snakes}
\usetikzlibrary{arrows.meta}
\tikzset{middlearrow/.style={
        decoration={markings,
            mark= at position 0.5 with {\arrow{#1}} ,
        },
        postaction={decorate}
    }
}

\makeatletter
 \AtBeginDocument{\let\LS@rot\@undefined}
\makeatother

\begin{document}
%\linenumbers
%\pagewiselinenumbers
%%%%%%%%%%%%%%%%%%%%%%%%%%%%%%%%%%%%%%%%%%%%%%%%%%%%%%%%%%%%%%%%%%%%%%%%%%%%%%%
%%%%                     Title and authors                                 %%%%
%%%%%%%%%%%%%%%%%%%%%%%%%%%%%%%%%%%%%%%%%%%%%%%%%%%%%%%%%%%%%%%%%%%%%%%%%%%%%%%
\title{Segregated quantum phases of dipolar bosonic mixtures 
       in two-dimensional optical lattices}

\author{Rukmani Bai}
\affiliation{Physical Research Laboratory,
             Ahmedabad - 380009, Gujarat,
             India}
\affiliation{Indian Institute of Technology Gandhinagar,
             Palaj, Gandhinagar - 382355, Gujarat,
             India}
\affiliation{Institute for Theoretical Physics III and
             Center for Integrated Quantum Science and Technology,\\
             University of Stuttgart, 70550 Stuttgart,
             Germany}
\author{Deepak Gaur}
\affiliation{Physical Research Laboratory,
             Ahmedabad - 380009, Gujarat,
             India}
\affiliation{Indian Institute of Technology Gandhinagar,
             Palaj, Gandhinagar - 382355, Gujarat,
             India}
\author{Hrushikesh Sable}
\affiliation{Physical Research Laboratory,
             Ahmedabad - 380009, Gujarat,
             India}
\affiliation{Indian Institute of Technology Gandhinagar,
             Palaj, Gandhinagar - 382355, Gujarat,
             India}
\author{Soumik Bandyopadhyay}
\affiliation{Physical Research Laboratory,
             Ahmedabad - 380009, Gujarat,
             India}
\affiliation{Indian Institute of Technology Gandhinagar,
             Palaj, Gandhinagar - 382355, Gujarat,
             India}
\author{K. Suthar}
\affiliation{Physical Research Laboratory,
             Ahmedabad - 380009, Gujarat,
             India}
\affiliation{Institute of Theoretical Physics, Jagiellonian University
             in Krak\'ow, \L{}ojasiewicza 11, 30-348 Krak\'ow, Poland}
\author{D. Angom}
\affiliation{Physical Research Laboratory,
             Ahmedabad - 380009, Gujarat,
             India}

\begin{abstract}
 We identify the quantum phases in a binary mixture of dipolar bosons in 
two-dimensional optical lattices. Our study is motivated by the recent 
experimental realization of binary dipolar condensate mixtures of 
Er-Dy [Phys. Rev. Lett. 121, 213601 (2018)]. We model the system by using the 
extended two-species Bose-Hubbard model and calculate the ground-state 
phase diagrams by using mean-field theory. For selected cases we also 
obtain analytical phase boundaries by using the site-decoupled mean-field 
theory. For comparison we also examine the phase diagram of two-species 
Bose-Hubbard model. Our results show that the quantum phases with the 
long-range intraspecies interaction phase separate with no phase ordering. The 
introduction of the long-range interspecies interaction modifies the quantum 
phases of the system. It leads to the emergence of phase-separated quantum 
phases with phase ordering. The transition from the phase-separated quantum 
phases without phase ordering to phase ordered ones breaks the inversion 
symmetry. 
\end{abstract}

%\pacs{03.75.Mn,03.75.Hh,67.60.Bc,67.85.Bc}

%% 03.75.Mn     Multicomponent condensates; spinor condensates 
%% 03.75.Hh     Static properties of condensates; thermodynamical, 
%%              statistical, and structural properties 
%% 67.60.Bc     Boson mixtures
%% 67.85.Bc     Static properties of condensates 

\maketitle

%%%%%%%%%%%%%%%%%%%%%%%%%%%%%%%%%%%%%%%%%%%%%%%%%%%%%%%%%%%%%%%%%%%%%%%%%%%%%%%
%%%%%%%%%                      Introduction                       %%%%%%%%%%%%%
%%%%%%%%%%%%%%%%%%%%%%%%%%%%%%%%%%%%%%%%%%%%%%%%%%%%%%%%%%%%%%%%%%%%%%%%%%%%%%%

\section{Introduction}

 The Bose-Hubbard model \cite{hubbard_63, fisher_89} describes the physics of 
ultracold bosonic atoms trapped in optical lattices \cite{jaksch_98}. 
The variation of the hopping term, equivalent to kinetic terms in continuum 
models, in the Bose-Hubbard model drives a quantum phase transition from the 
Mott insulator (MI) to the superfluid (SF) phase. And this transition 
has been experimentally observed \cite{greiner_02}. The interparticle 
interaction in the Bose-Hubbard model is onsite or contact in nature. The 
introduction of the nearest neighbor (NN) interaction in the Bose-Hubbard 
model generates two more phases: density wave (DW) and supersolid (SS).
This model with the NN interactions is referred to as the extended
Bose-Hubbard model \cite{kuhner_00} and shows rich physics compared with the 
Bose-Hubbard model. Such a model captures the physics of dipolar ultracold
quantum gases in optical lattices~\cite{baier_16,bandyopadhyay_19}. A more 
complex system, ideal to model several condensed-matter systems, is to fill 
the optical lattice with two species Bose-Einstein Condensate or binary 
condensate. A binary condensate could be a condensate mixture of two different 
atomic species \cite{modugno_02,lercher_11,mccarron_11,pasquiou_13,wacker_15,
wang_16}, two hyperfine states of an atom
\cite{myatt_97,hall_98,stamper_98,stenger_98,maddaloni_00,delannoy_01,
sadler_06,mertes_07,anderson_09,tojo_10} or two different isotopes of an 
atomic species \cite{papp_08,handel_11,sugawa_11a}. It was experimentally 
first realized in the two hyperfine states $|F = 2, m_F = 2 \rangle$ and 
$|F = 1, m_F = -1 \rangle$ of $^{87}$Rb atom \cite{myatt_97}. The binary
condensates, in the weakly interacting continuum systems, have been used to 
investigate novel phenomena such as pattern formation 
\cite{sasaki_09,gautam_10_1,gautam_10_2,ronen_08,hoefer_11,hamner_11,de_14}, 
phase separation \cite{ho_96,ao_98,gautam_11,roy_15_2,bandyopadhyay_17,papp_08,
tojo_10,mccarron_11,wacker_15,wang_16}, nonlinear dynamical excitations 
\cite{gautam_12,gautam_13,roy_14_2,kuopanportti_19,mertes_07,eto_16a,eto_16}, 
collective excitations \cite{roy_14_1,suthar_15,roy_15_1,suthar_16,
suthar_17,pal_17,pal_18}, Kibble-Zurek mechanism \cite{nicklas_15}, and the 
production of dipolar molecules \cite{molony_14,guo_16,will_16}. The phase 
separation, among all the phenomena is a unique property of binary 
condensates. In this work we study the binary condensates trapped in the 
optical lattices that can be described by the Bose-Hubbard model with 
appropriate modifications. The experimental realization in optical lattices 
are reported in Refs. \cite{catani_08, gadway_10} and early theoretical studies
are presented in Refs. \cite{altman_03,chen_03,kuklov_03, kuklov_04}.
A remarkable recent achievement related to binary condensates is the 
experimental realization with dipolar quantum mixtures of Er-Dy, reported 
in a recent work \cite{trautmann_18}.

The physics of the two-species Bose-Hubbard model (TBHM), the lattice
counterpart of a binary condensate, in one dimension has been investigated in
detail \cite{ mishra_07, zhan_14,wang_14}. And, there has been some works in
two dimension as well \cite{altman_03,chen_03, kuklov_03,kuklov_04,
isacsson_05,iskin_10,anufriiev_16,boninsegni_01}. The phase diagram of TBHM 
shows different combinations of mixed MI-SF phases apart from the Mott 
insulator and superfluid phases. And, these have been investigated by using 
quantum Monte Carlo \cite{kuklov_03, kuklov_04}, mapping to spin systems 
\cite{altman_03}, and with mean-field theory \cite{chen_03,isacsson_05,
iskin_10,pai_12,anufriiev_16}. These studies, except for Ref. \cite{pai_12}, 
considered homogeneous systems. However, hitherto the phenomenon of phase 
separation in two-dimensional TBHM has yet to be investigated in detail. 

The quantum phases of TBHM in the phase-separated domain, unlike in the
binary condensates, do not show segregation into two spatial domains. We 
attribute this to the lack of long-range interactions. The simplest 
modification to include the effect of long-range interactions is to add 
nearest neighbor interactions. The extended Bose-Hubbard model, as mentioned 
earlier, supports two more quantum phases: density wave  
\cite{capogrosso_10,flottat_17,iskin_11} and supersolid 
\cite{ng_08,iskin_11,yamamoto_11,boninsegni_12,leonard_17,kuno_14}.
The density wave phase is an insulating phase similar to the Mott insulator 
phase but it has crystalline order or diagonal long-range order. And, the 
supersolid phase is a compressible phase with both diagonal and off-diagonal 
long-range order. In a recent study of extended TBHM (eTBHM)~\cite{wilson_16}, 
it was shown that the supersolid phase exists for  small value of 
NN interactions. In this work, the NN interaction was limited to either one of 
the species or between the species. We address this research gap by 
including all the possible intra- and interspecies NN interactions. Such a 
model is apt to describe the physics of dipolar Bose-Bose mixtures in optical 
lattices. An example of such a combination is the recently realized 
Er-Dy mixture \cite{trautmann_18}. An important result of our work is the 
possibility to realize compressible and incompressible quantum phases with 
spatial segregation. Such a phase could be instrumental in examining 
superfluid instabilities and other nonequilibrium properties in the lattice 
models of quantum liquids.

The remainder of the paper is organized into four sections. In Sec. 
\ref{sec_theory} we describe the zero-temperature Hamiltonian of the TBHM 
and discuss the Gutzwiller mean-field theory of the model. We then discuss 
the mean-field decoupling theory to calculate the 
compressible-incompressible phase boundaries analytically. This is followed 
by a brief discussion on the characterization of quantum phases. The phase 
diagrams of TBHM are discussed in the Sec. \ref{sec_phdiag}. Section 
\ref{sec_phdiag_ext} includes a discussion on the phase diagram of the eTBHM.
In particular, the miscible and immiscible phases. We also check the dynamical
stability of the quantum phases by computing the collective excitations of the
system. We end the paper with conclusion in Sec. \ref{sec_conclude}.

%%%%%%%%%%%%%%%%%%%%%%%%%%%%%%%%%%%%%%%%%%%%%%%%%%%%%%%%%%%%%%%%%%%%%%%%%%%%%
%%%%%%  Section: BH Hamiltonian for TBEC
%%%%%%%%%%%%%%%%%%%%%%%%%%%%%%%%%%%%%%%%%%%%%%%%%%%%%%%%%%%%%%%%%%%%%%%%%%%%%

\section{Theory} \label{sec_theory}
\subsection{Two-species Bose-Hubbard model Hamiltonian}
  At zero temperature, the TBHM Hamiltonian, which describes the physics of a 
binary condensate in a two-dimensional optical lattice, is \cite{damski_03}
\begin{eqnarray}
\hat{H}^{{\rm TBH}} &=& -\sum_{p, q, k}\bigg [ \Big( J_x^k 
              \hat{b}_{p+1, q}^{\dagger k}\hat{b}_{p, q}^k + {\rm H.c.}\Big)
              + \Big( J_y^k \hat{b}_{p, q+1}^{\dagger k}
              \hat{b}_{p, q}^k  \nonumber\\ 
              &&+ {\rm H.c.}\Big) - \frac{U_{kk}}{2}
              \hat{n}_{p, q}^k (\hat{n}_{p, q}^k-1) + 
              \tilde{\mu}^k_{p,q}\hat{n}_{p, q}^k\bigg]
              \nonumber\\
              &&+\sum_{p, q} U_{12}\hat{n}_{p, q}^1 \hat{n}_{p, q}^2,
\label{tbhm}  
\end{eqnarray}
where $k = 1$,$2$ is the species index, $(p,q)$ are the lattice indices, 
$J_x^k$ ($J_y^k$) is the NN hopping strength along $x$ ($y$) directions, 
$\hat{b}^{\dagger k}_{p,q}$ ($\hat{b}^k_{p,q}$) is the creation (annihilation)
operator, and $\hat{n}_{p, q}^k$ is the number operator at site ($p,q$). 
$U_{kk}$ is intraspecies interaction strength, and $U_{12}$ is the interspecies 
interaction strength between two species. Furthermore, 
$\tilde{\mu}^k_{p,q} =\mu^k - \varepsilon_{p,q}^k$, is the local chemical 
potential at each site for the two species where $\varepsilon_{p,q}^k$ is 
the envelop potential for the species. For a system of $K \times L$ lattices 
sites, the index along $x$ ($y$) has values $p=1,\ldots K$ ($q=1,\ldots L$). 
The unique feature of the binary condensates is the phase separation and for 
continuum systems, the criterion for phase segregation is 
$U_{12}^2 > U_{11}U_{22}$ \cite{ho_96, Trippenbach_00}. Otherwise, it is in 
the miscible phase. For the case of strongly interacting binary condensates in
optical lattices, described by the above Hamiltonian, we show the existence of
different phases in both the miscible and immiscible domains.
 
To obtain the ground state of the Hamiltonian in Eq.~(\ref{tbhm}), we use 
single-site Gutzwiller mean-field (SGMF) theory 
\cite{rokhsar_91, sheshadri_93,bai_18, pal_19, bandyopadhyay_19, suthar_20}.
The starting point of this theory is to separate the operators into mean-field
and fluctuation operator components as 
$\hat{b}_{p, q}^k = \phi_{p,q}^k + \delta \hat{b}_{p, q}^k\;$ and 
$\;\hat{b}^{\dagger k}_{p,q} = \phi_{p,q}^{k*} + 
\delta\hat{b}_{p,q}^{\dagger k}$. Then, the Hamiltonian in Eq. (\ref{tbhm}) is 
reduced to the sum of the single-site mean-field Hamiltonian
\begin{eqnarray}
\hat{h}_{p,q}^{{\rm TBH}} &= &- \sum_{k}\left[J_x^k 
                  \left(\hat{b}_{p+1, q}^{\dagger k}\phi_{p,q}^k 
                  + \phi^{k*}_{p + 1, q}\hat{b}_{p, q}^k \right ) + {\rm H.c.}
                  \right.  \nonumber\\
                  && +  J_y^k\left(\hat{b}_{p, q+1}^{\dagger k} \phi_{p,q}^k 
                  +  \phi^{k*}_{p, q+1}\hat{b}_{p, q}^k  \right) + {\rm H.c.}
                  \nonumber \\
                  &&-\left . \frac{U_{kk}}{2}\hat{n}_{p, q}^k
                  \left (\hat{n}_{p, q}^k-1\right) 
                  + \tilde{\mu}^k_{p,q}\hat{n}_{p, q}^k\right]
                  + U_{12}\hat{n}_{p, q}^1 \hat{n}_{p, q}^2,
                  \nonumber\\
\label{ham_ss_tbec}
\end{eqnarray}
where $\phi_{p,q}^k$ ($\phi_{p,q}^{k*}$) is the superfluid order parameter. 
With this definition of the single-site mean-field Hamiltonian, the
total Hamiltonian of the system is 
\begin{equation}
   \hat{H}^{{\rm TBH}} = \sum_{p,q}\hat{h}_{p,q}^{{\rm TBH}}.
\end{equation}
For the details of the derivations, see Ref. \cite{bai_18}. To get the ground
state we diagonalize the Hamiltonian in Eq.~(\ref{ham_ss_tbec}) at each site.
And, for this we use the Gutzwiller ansatz, based on which the ground state
at site ($p,q$) is \cite{anufriiev_16}
\begin{equation}
  |\psi\rangle_{p,q} = \sum_{n_1, n_2 }c^{(p,q)}_{n_1, n_2} 
                       |n_1, n_2\rangle_{p,q}.
  \label{gw_2s}
\end{equation}
Here, $|n_1, n_2\rangle$ is a Fock state, which is the direct product of the 
$n_1$ and $n_2$ occupation number states of the first and second species, 
respectively. The occupation number states $n_k\in [0,N_b -1]$, 
where $N_b$ is the total number of local Fock states used in the computation,
and $c_{n_1, n_2}^{p,q}$ are complex co-efficients with 
$\sum_{n_1,n_2} |c^{(p,q)}_{n_1,n_2}|^2$ = 1. From the ground state, we can
compute the new superfluid order parameter of the two species as
\begin{subequations}
  \begin{eqnarray}
    \phi_{p, q}^1& =& _{p,q}\langle\psi|\hat{b}_{p, q}^1|\psi\rangle_{p,q} 
            = \sum_{n_1, n_2}\sqrt{n_1} 
              {c^{(p,q)*}_{n_1-1, n_2}}c^{(p,q)}_{n_1,n_2},
              \label{gw_phi2s_1}\\
    \phi_{p, q}^2& =& _{p,q}\langle\psi|\hat{b}_{p, q}^2|\psi\rangle_{p,q} 
            = \sum_{n_1, n_2}\sqrt{n_2} 
              {c^{(p,q)*}_{n_1, n_2-1}}c^{(p,q)}_{n_1,n_2}.
    \label{gw_phi2s_2}              
  \end{eqnarray}
\end{subequations}

Similarly, corresponding lattice occupancies are
\begin{subequations}
  \begin{eqnarray}
   \rho_{p, q}^1& =& _{p,q}\langle\psi|\hat{n}_{p, q}^1|\psi\rangle_{p,q} 
            = \sum_{n_1, n_2} n_1 |c^{(p,q)}_{n_1,n_2}|^2,
            \label{num2s_1}\\
   \rho_{p, q}^2& =& _{p,q}\langle\psi|\hat{n}_{p, q}^2|\psi\rangle_{p,q} 
            = \sum_{n_1, n_2}n_2|c^{(p,q)}_{n_1,n_2}|^2. 
   \label{num2s_2}              
  \end{eqnarray}
\end{subequations}
Using the new superfluid order parameters, the ground state of the next 
lattice site is computed and this process is repeated until all the 
lattices sites are covered. One such sweep is identified as an 
iteration and we then, start the process again for the next iteration.
The iterations are carried out until the convergence criterion 
$|\phi_{p,q}^{n-1} - \phi_{p,q}^n| \lesssim 10^{-12}$ is satisfied at the 
$n^{{\rm th}}$ iteration. In the present work, to determine the phase 
diagrams, we consider lattice system of size $10 \times 10$ and choose 
$N_b = 7$. That is, $K$ and $L$ are both 10. We find that the phase 
boundaries remain unchanged when the system size is augmented to 
$20\times 20$. We also use the augmented system size to validate key 
findings. In addition, we employ periodic boundary conditions to model an 
infinite-sized system.

%%%%%%%%%%%%%%%%%%%%%%%%%%%%%%%%%%%%%%%%%%%%%%%%%%%%%%%%%%%%%%%%%%%%%%%%%%%%
%%%% Subsection: Extended TBHM    
%%%%%%%%%%%%%%%%%%%%%%%%%%%%%%%%%%%%%%%%%%%%%%%%%%%%%%%%%%%%%%%%%%%%%%%%%%%%

\subsection{Extended two-species Bose-Hubbard model Hamiltonian}
  The Bose-Hubbard model with NN interaction, referred to as the extended 
Bose-Hubbard model, exhibits a richer phase diagram than does the Bose-Hubbard
model and it has the novel feature of harbouring the supersolid phase. The
phase diagram of this model consists of density wave, supersolid, Mott
insulator and superfluid phases. Similarly, the eTBHM also exhibits these
phases as well as miscible and segregated phases and the model Hamiltonian of
the system is
\begin{eqnarray}
\!\!\!\!\hat{H}^{{\rm ext}} &=& \hat{H}^{{\rm TBH}} 
                       + \sum_{p, q, k} \biggl [ V_k \hat{n}^k_{p,q}  
                         \Big ( \hat{n}^k_{p-1,q} + \hat{n}^k_{p+1,q} 
                       + \hat{n}^k_{p,q -1}  \nonumber\\
                    && + \hat{n}^k_{p,q+1}\Big )+  V_{12} \hat{n}_{p,q}^k
                         \Big(\hat{n}^{3-k}_{p-1,q} + \hat{n}^{3-k}_{p+1,q} 
                       + \hat{n}^{3-k}_{p,q -1} \nonumber \\
                    && + \hat{n}^{3-k}_{p,q+1}\Big ) \biggr],
\label{bhm_ext}
\end{eqnarray}
here $V_k$ and $V_{12}$ are the intraspecies and interspecies NN interaction 
strengths respectively. In the experiments the ratio of NN interaction to the
on-site interaction can be varied by tuning the on-site interaction through a 
magnetic Feshbach resonance. The NN interaction, arising from the dipole-dipole 
interaction, can also be varied by rotation of the dipoles with a 
time-dependent external magnetic field ~\cite{tang_18,giovanazzi_02}. Using 
this method it can even be turned off. The quantum phases obtained from the 
model described by the above Hamiltonian are relevant to the experimental 
realizations with the dipoles oriented perpendicular to the lattice plane. In a 
latter section, Section \ref{fint_tilt_ang}, we provide a brief description of 
the quantum phases when the tilt angle $\theta$ is nonzero. Here, $\theta$ is 
the angle between the orientation of the dipoles and the normal to the lattice 
plane. Thus, to relate with the experimental observations and
predict possible phases we vary the inter- and intraspecies interaction
strengths. We use SGMF theory to obtain the ground state of the system,
then, in this method the total Hamiltonian is the sum of the single-site
mean-field Hamiltonian
\begin{eqnarray}
\!\!\hat{h}_{p,q}^{{\rm ext}} &=& \hat{h}_{p,q}^{{\rm TBH}}  
                            + \sum_{k} \biggl [ V_k \hat{n}^k_{p,q}  
                               \Big ( \langle\hat{n}^k_{p-1,q}\rangle 
			       + \langle \hat{n}^k_{p+1,q} \rangle
                            + \langle\hat{n}^k_{p,q -1}\rangle \nonumber \\
			   &&+ \langle \hat{n}^k_{p,q+1} \rangle\Big ) 
			    +  V_{12} \hat{n}_{p,q}^k 
                              \Big(\langle \hat{n}^{3-k}_{p-1,q}\rangle
			   + \langle \hat{n}^{3-k}_{p+1,q} \rangle
			   + \langle \hat{n}^{3-k}_{p,q -1} \rangle \nonumber\\ 
			   &&+ \langle\hat{n}^{3-k}_{p,q+1}\rangle\Big )\biggr].
\label{bhm_ext_ss}
\end{eqnarray}
We diagonalize this Hamiltonian at each site separately, and
obtain the ground state. The NN-interaction term contributes to the diagonal
matrix element. From the single-site wavefunction, the superfluid order 
parameter and lattice occupancy can be calculated from the expressions in
Eqns.~(\ref{gw_phi2s_1}),(\ref{gw_phi2s_2})
and~(\ref{num2s_1}), (\ref{num2s_2}).

%%%%%%%%%%%%%%%%%%%%%%%%%%%%%%%%%%%%%%%%%%%%%%%%%%%%%%%%%%%%%%%%%%%%%%%%%%%%
%%%% Subsection: Mean-field decoupling theory
%%%%%%%%%%%%%%%%%%%%%%%%%%%%%%%%%%%%%%%%%%%%%%%%%%%%%%%%%%%%%%%%%%%%%%%%%%%%

\subsection{Mean-field decoupling theory}

%%%%%%%%%%%%%%%%%%%%%%%%%%%%%%%%%%%%%%%%%%%%%%%%%%%%%%%%%%%%%%%%%%%%%%%%%%%%
%%%%  Subsubsection: TBHM
%%%%%%%%%%%%%%%%%%%%%%%%%%%%%%%%%%%%%%%%%%%%%%%%%%%%%%%%%%%%%%%%%%%%%%%%%%%%

\subsubsection{Two-species Bose-Hubbard model} 
 To calculate the phase boundaries between Mott insulator and superfluid 
phases analytically we use the site decoupled mean-field 
theory~\cite{oosten_01,iskin_09, bandyopadhyay_19}. 
For this, we adapt perturbative analysis of the mean-field Hamiltonian in 
Eq.~(\ref{ham_ss_tbec}). It is important to note that the superfluid order 
parameter $\phi^{k}_{p,q} $ is zero in the Mott insulator phase, but nonzero 
in the superfluid phase. So, the vanishing of the superfluid order parameter 
$\phi^{k}_{p,q} \rightarrow 0^{+}$ marks the MI-SF phase boundary in the phase 
diagram. With this consideration, in the site-decoupled mean-field theory, 
the interaction and the chemical potential terms constitute the unperturbed 
Hamiltonian $\hat{h}_{p,q,0}^{\rm TBH}$. From Eq. (\ref{ham_ss_tbec}),
\begin{eqnarray}
\hat{h}_{p,q,0}^{{\rm TBH}} 
	        &= & \sum_{k}\left[ \frac{U_{kk}}{2}\hat{n}_{p, q}^k
                  \left (\hat{n}_{p, q}^k-1\right) 
                  - \tilde{\mu}^k_{p,q}\hat{n}_{p, q}^k\right]
                  \nonumber\\
                  &&+ U_{12}\hat{n}_{p, q}^1 \hat{n}_{p, q}^2,
\label{ham_ss_h0}
\end{eqnarray}
which is diagonal with respect to the Fock basis states. Then, the hopping
terms in Eq.~(\ref{ham_ss_tbec}) act as the perturbation,
\begin{eqnarray}
\hat{h}_{p,q,1}^{{\rm TBH}} &= &- \sum_{k}\Bigl [J_x^k 
                  \left(\hat{b}_{p+1, q}^{\dagger k}\phi_{p,q}^k 
                  + \phi^{k*}_{p + 1, q}\hat{b}_{p, q}^k \right ) + {\rm H.c.}
                   \nonumber\\
                  && +  J_y^k\left(\hat{b}_{p, q+1}^{\dagger k} \phi_{p,q}^k 
                  +  \phi^{k*}_{p, q+1}\hat{b}_{p, q}^k  \right) + {\rm H.c.}
                   \Bigr ],
\label{ham_ss_h1}
\end{eqnarray}
with the superfluid order parameter $\phi^{k}_{p,q}$ as the perturbation
parameter. Then, from the first-order perturbative correction to the
ground-state wavefunction (details given in Appendix~\ref{appendix_a}), we have 
\begin{eqnarray}
\!\!\!\!\!\!\!\!\!\phi^{k}_{p,q} = J\bar{\phi}^{k}_{p,q} 
	\left(\frac{n^{k}_{p,q} +1}{n^{k}_{p,q}U -\bar{\mu}^{k}_{p,q}}
		-\frac{n^k_{p,q}}{(n^k_{p,q} -1)U -\bar{\mu}^{k}_{p,q}}\right),
 \label{order_par_ana}
\end{eqnarray}
with
\begin{eqnarray}
   \bar{\mu}^{k}_{p,q} &= &\tilde{\mu}^{k}_{p,q}-U_{12}n^{3-k}_{p,q},
                          \nonumber \\
   \bar{\phi}^{k}_{p,q} &= &\left(\phi^{k}_{p+1,q}+\phi^{k}_{p-1,q}
                           +\phi^{k}_{p,q+1}+\phi^{k}_{p,q-1}\right).
                           \nonumber
\end{eqnarray}
For a homogeneous lattice system $\varepsilon^{k}_{p,q} = 0$. Then, in the 
Mott insulator phase the total density $\rho = \rho^{1}+\rho^{2}$ is integer 
commensurate and $\phi^{k}_{p,q}=0$. In the superfluid phase, the order 
parameter is nonzero and uniform, say $\phi^{k}_{p,q}=\varphi^{k}_{0}$. With 
these considerations, 
$\bar{\phi}^{k}_{p,q} = \bar{\phi}^{k} = 4\varphi^{k}_{0}$. Starting from the
superfluid phase, at the SF-MI phase boundary 
$\varphi^{k}_{0}\rightarrow 0^{+}$. 
Considering this limit in Eq.~(\ref{order_par_ana}), we obtain the 
equation which defines the phase boundary in terms of $J$ for a 
particular value of $\mu$. \\

For the $\rho = 2$ Mott lobe, in the miscible domain, atoms of both the species
fill all the lattice sites. That is, $n^1_{p,q} =n^2_{p,q} =1$. The MI-SF phase
boundary is, then, defined by
\begin{eqnarray}
   \frac{1}{4J}= \frac{2}{U -\mu +U_{12}} +\,\frac{1}{\mu -U_{12}}.
   \label{tbhm_even}
\end{eqnarray}
On the other hand for finite $U_{12}$, the system is in the immiscible domain
for the $\rho=1$ Mott lobe. The density pattern has one atom at each lattice 
site chosen randomly from the two species. Thus, at a given lattice site 
$(p,q)$ we can have the occupancies as $n^{1}_{p,q} =1, n^{2}_{p,q} =0$ 
or $n^{1}_{p,q} =0, n^{2}_{p,q} =1$. In the perturbative analysis, without
loss of generality, we consider neighboring lattice sites which are occupied
by atoms of different species. This is also one realization of the
energetically favourable configuration for $U_{12}<U$. Then, with the
correction arising from $b^{\dagger 1}\phi^{1}$, the equation
\begin{eqnarray}
   \frac{1}{4J}= \frac{2}{U_{12} -\mu} +\,\frac{1}{\mu},
   \label{tbhm_odd}
\end{eqnarray}
defines the phase boundary of the Mott lobe with $\rho=1$. Based on similar 
analysis, we can obtain the phase boundary of other Mott lobes. For which 
we have to use Eqs.(\ref{tbhm_even}) and (\ref{tbhm_odd}) for the even and 
odd-integer values of $\rho$, respectively.

%%%%%%%%%%%%%%%%%%%%%%%%%%%%%%%%%%%%%%%%%%%%%%%%%%%%%%%%%%%%%%%%%%%%%%%%%%%%
%%%%  Subsubsection: Extended TBHM
%%%%%%%%%%%%%%%%%%%%%%%%%%%%%%%%%%%%%%%%%%%%%%%%%%%%%%%%%%%%%%%%%%%%%%%%%%%%

\subsubsection{Extended two-species Bose-Hubbard model}
 We extend the analysis done in previous section to the eTBHM case. 
The expression of the order parameter is similar to Eq.~(\ref{order_par_ana}) 
but $\bar{\mu}^{k}_{p,q}$ is given by
\begin{eqnarray}
   \bar{\mu}^{k}_{p,q} &= &\tilde{\mu}^{k}_{p,q}-U_{12}n^{3-k}_{p,q}
		-4V_k n^{k}_{p,q} -4V_{12} n^{3-k}_{p,q},
\label{mu_bar_etbhm}
\end{eqnarray} 
For the MI(1,1) phase with $\rho=2$, the occupancies are $n^{k}_{p,q}=1$.
Furthermore, assuming $V_1 =V_2$, the MI-SF boundary is given by
\begin{eqnarray}
   \frac{1}{4J}= \frac{2}{U -\bar{\mu}} +\,\frac{1}{\bar{\mu}}.
 \label{exTBHM_MI2}
\end{eqnarray}
with $\bar{\mu} = \mu -U_{12} -4V_1 -4V_{12}$. Similarly, the 
phase boundary for the MI(2, 2) lobe can be obtained by choosing
$n^{k}_{p,q} =2$ in Eq.(\ref{order_par_ana}) with $\bar{\mu}^{k}_{p,q}$ given
by Eq.(\ref{mu_bar_etbhm}). In the density wave phase, the two 
sublattice structure description is applicable. Using this, the 
density wave to compressible phase boundary 
for $V_1=V_2$ and $V_{12} =0$ is given by
\begin{eqnarray}
\!\!\!\! \frac{1}{16J^2}&=& \left\{\frac{n^{1,B} +1}
             {Un^{1,B} -\mu +U_{12}n^{2,B} +4V_1 n^{1,A}} \right. \nonumber\\
             &&\left. - \frac{n^{1,B}}
             {U(n^{1,B}-1) -\mu +U_{12}n^{2,B} +4V_1 n^{1,A}} 
             \right\}\nonumber \\ 
             &&\times \left\{\frac{n^{1,A} +1}
             {Un^{1,A} -\mu +U_{12}n^{2,A} +4V_1 n^{1,B}} \right. \nonumber\\
             &&\left. - \frac{n^{1,A}}
             {U(n^{1,A}-1) -\mu +U_{12}n^{2,A} +4V_1 n^{1,B}}
             \right\}\nonumber \\
\label{order_par_ana_etbhm}
\end{eqnarray}
The details are given in Appendix \ref{appendix_b}. As an example consider
the DW(1,0) phase. It has $n^{1,A}=1$, $n^{1,B}=0$, $n^{2,A}=0$ and 
$ n^{2,B}=1$. From the above equation, the DW(1,0)-compressible phase 
boundary is given by
\begin{eqnarray}
\frac{1}{16J^2} = \left\{ \frac{1}{-\mu +U_{12} +4V_1} \right\}
                      \times \left\{ \frac{2}{U -\mu} + \frac{1}{\mu} \right\}.
\end{eqnarray}
Using Eq. (\ref{order_par_ana_etbhm}), we can also calculate the 
phase boundaries for other density wave phases.

%%%%%%%%%%%%%%%%%%%%%%%%%%%%%%%%%%%%%%%%%%%%%%%%%%%%%%%%%%%%%%%%%%%%%%%%%%%%
%%%% Subsection: Characterization of the phases
%%%%%%%%%%%%%%%%%%%%%%%%%%%%%%%%%%%%%%%%%%%%%%%%%%%%%%%%%%%%%%%%%%%%%%%%%%%%

\subsection{Characterization of the phases}
\begin{table}[h!]
  \begin{ruledtabular}
 \begin{tabular}{lllll}
       \text{Quantum phase} &$\rho$ &$\phi$ &$\Delta\rho^k$ &$\Delta\phi^k$\\
      \colrule
 Mott Insulator & \text{Integer} & $ = 0$   & $ = 0$   & $ = 0$\\
 Superfluid     & \text{Real}    & $ \ne 0$ & $ = 0$   & $ = 0$\\
 Density Wave   & \text{Integer} & $ = 0$   & $ \ne 0$ & $ = 0$\\
 Supersolid     & \text{Real}    & $ \ne 0$ & $ \ne 0$ & $ \ne 0$\\
    \end{tabular}
    \caption{Classification of different quantum phases with
    order parameters for our systems.}
      \label{table}
  \end{ruledtabular}
\end{table}

To identify different quantum phases of the system we compute the
density contrast $\Delta\rho^k$, order parameter contrast $\Delta\phi^k$
and compressibility $\kappa^k$. To define $\Delta\rho^k$, divide the lattice 
site occupancies as
\begin{equation}
   n^k_{p,q}=\begin{cases}
                n^{k,A} & \text{if $(p,q) \in$ sublattice A} \\
                n^{k,B} & \text{if $(p,q) \in$ sublattice B},
             \end{cases}
   \label{occ_def}
\end{equation}
then, the density contrast of the $k$th species is
\begin{equation}
   \Delta\rho^k = n^{k,A}-n^{k,B}.
   \label{den_con}
\end{equation}
The order parameter contrast is defined similarly as
\begin{equation}
   \Delta\phi^k = \phi^{k,A}-\phi^{k,B},
   \label{den_con}
\end{equation}
where $\phi^{k,A}$ and $\phi^{k,B}$, like in the case of density are the 
values of the order parameters at lattice sites with $(p,q)$ belonging to 
sublattices A and B, respectively. The compressibility of each species are 
calculated by using the definition $\partial \mu^k/\partial \rho^k$.

 The TBHM, like the single species Bose-Hubbard model, shows two phases, 
Mott insulator and superfluid. The Mott insulator phase is an incompressible 
phase with integer commensurate density $n^{k,A}=n^{k,B}\in\mathbb{N}$. And 
incompressibility implies zero superfluid order parameter 
$\phi^{k,A}=\phi^{k,B}=0$. The superfluid phase, on the other
hand is compressible. Hence, it has $n^{k,A}=n^{k,B}\in \mathbb{R}$, 
$\phi^{k,A}=\phi^{k,B}\in \mathbb{R}$ and $\kappa^k\neq0$. For these two 
phases, the density and superfluid order parameters are uniform, 
so the contrast order parameters $\Delta\rho^k$ and $\Delta\phi^k$ are zero. 
In the eTBHM, the NN interaction leads to the emergence of two more quantum 
phases, density wave and supersolid. These two phases have nonuniform density 
and superfluid order parameters. As a result
the distinguishing features of these phases are nonzero contrast order
parameters. The density wave phase has integer $n^k_{p,q}$ with 
$n^{k,A}\neq n^{k,B}$
and $\Delta \rho^k\in\mathbb{N}$. This phase has zero superfluid order parameter
$\phi^{k,A}=\phi^{k,B}=0$ and hence, is incompressible. The supersolid phase
has real $n^k_{p,q}$ with $n^{k,A}\neq n^{k,B}$ and
$\Delta \rho^k\in\mathbb{R}$. 
The superfluid order parameter in this phase is nonzero and nonuniform. 
This implies that $n^{k,A}\neq n^{k,B}$ and $\phi^{k,A}\neq\phi^{k,B}$. So, 
both the contrast order parameters are nonzero in this phase. For easy 
reference the properties of the different quantum phases are listed in 
Table. \ref{table}.

%%%%%%%%%%%%%%%%%%%%%%%%%%%%%%%%%%%%%%%%%%%%%%%%%%%%%%%%%%%%%%%%%%%%%%%%%%%%%%%
%%%% Section: Phase diagram of TBHM
%%%%%%%%%%%%%%%%%%%%%%%%%%%%%%%%%%%%%%%%%%%%%%%%%%%%%%%%%%%%%%%%%%%%%%%%%%%%%%%

\section{Phase diagram of two-species Bose-Hubbard model} \label{sec_phdiag}
\begin{figure}[t]
%  \begin{center}
  \includegraphics[width=8.0cm]{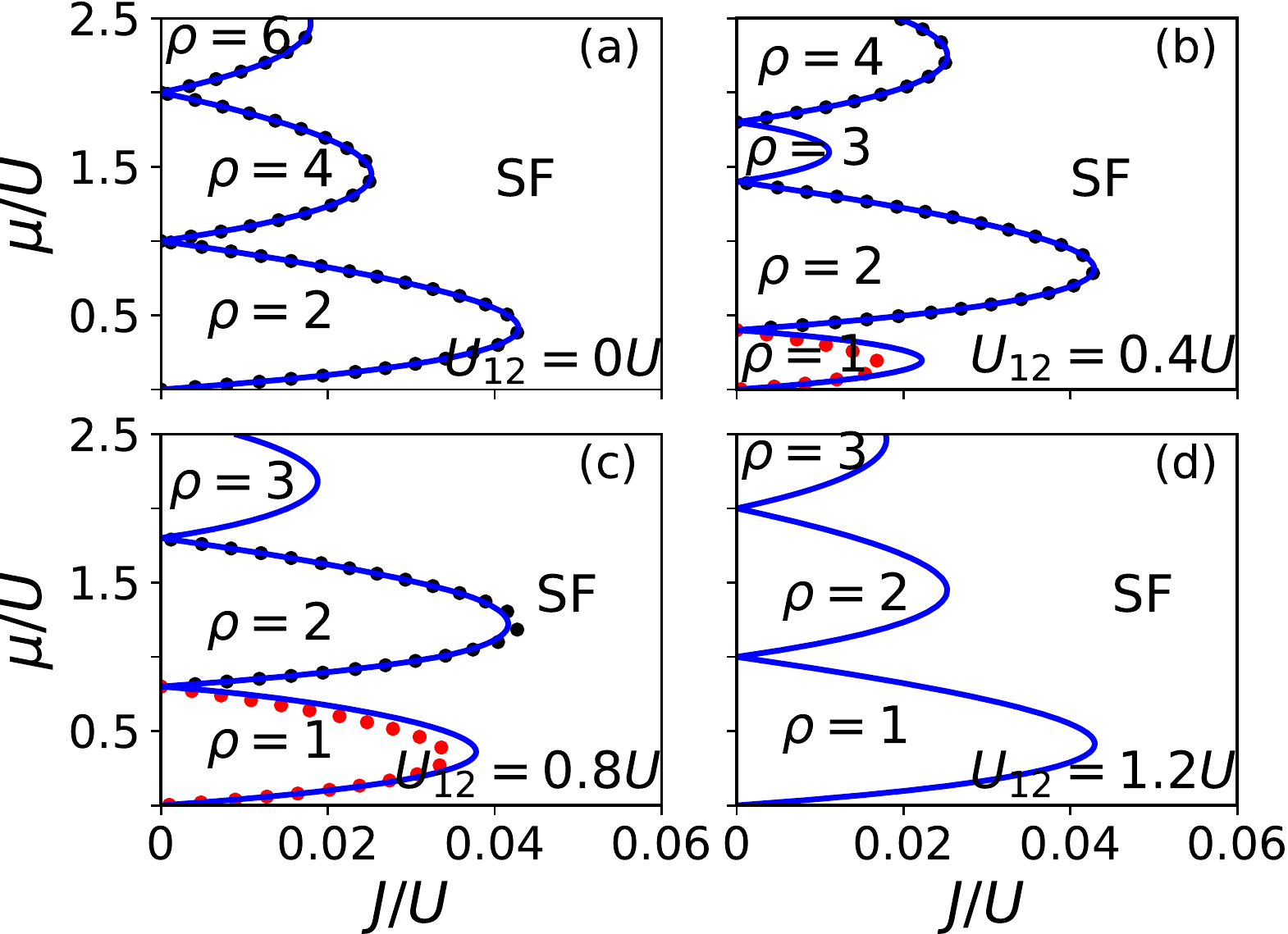}
  \caption{Phase diagram of TBHM by varying the interspecies interaction
           strength $U_{12}$. Blue solid lines represent numerically obtained
           phase boundaries for the mean field Hamiltonian. Filled dots marks
           phase boundaries between compressible and
           incompressible phases, obtained analytically by perturbative
           analysis of the mean-field Hamiltonian. The odd occupancy Mott lobes
           appear for nonzero $U_{12}$ and enlarges with increasing $U_{12}$.}
  \label{tbec_ph}
%  \end{center}
\end{figure}

To compute the ground-state wavefunction and determine the phase, we 
initialize the superfluid order parameter  $\phi$.  This, then, defines 
the Hamiltonian in Eq.~(\ref{ham_ss_tbec}) and Hamiltonian matrix elements are 
computed by using the Gutzwiller wavefunction in Eq.~(\ref{gw_2s}). By
diagonalizing the Hamiltonian matrix for each site we obtain the ground-state
wavefunction. From the results, the MI-SF phase boundary is identified 
based on the superfluid order parameter and the lattice occupancy. 
For the incompressible Mott insulator phase, at each lattice site,
$\phi$ is zero and $\rho$ is integer commensurate. For the superfluid phase,  
$\phi$ is nonzero and $\rho$ is real commensurate. The phase diagrams of 
TBHM given in Eq.~(\ref{ham_ss_tbec}) for different values of $U_{12}$ are 
shown in Fig.~\ref{tbec_ph}. 

For simplicity, we consider symmetric hopping $J_x^k = J_y^k = J$, equal 
chemical potential $\tilde{\mu}^1_{p,q} = \tilde{\mu}^2_{p,q} = \mu$ and 
identical intraspecies interactions $U_{kk} = U$. We scale all the energies
with $U$, and define the phase diagram in the $J/U-\mu/U$ plane.

%%%%%%%%%%%%%%%%%%%%%%%%%%%%%%%%%%%%%%%%%%%%%%%%%%%%%%%%%%%%%%%%%%%%%%%%%%%%
%%%% subsection: zero temperature phase diagram
%%%%%%%%%%%%%%%%%%%%%%%%%%%%%%%%%%%%%%%%%%%%%%%%%%%%%%%%%%%%%%%%%%%%%%%%%%%%

\subsection{Zero temperature phase diagram}
The phase diagram consists of a sequence of Mott lobes having integer $\rho$. 
Without the interspecies interaction $U_{12} = 0$, as shown in
Fig. \ref{tbec_ph}(a), the phase diagram is equivalent to the case of single 
species, but with twice the occupancy. That is the Mott lobes, which have 
$\rho=2n$ with $\rho^1=\rho^2=n$ and $n\in\mathbb{N}$. So, the lowest Mott lobe
has $\rho=2$ and each lattice has one atom from each of the two species. As a 
result, the phase diagram is identical to the single species case. With the 
introduction of the interspecies interaction ($U_{12} \neq 0$) the half filled
lobes like $\rho = 1$ emerge in the phase diagram with $0 < \rho^1 <1$, and 
then, $\rho^2 = 1-\rho^1$. This is discernible for $U_{12} =0.4 U$ from the
Fig.~\ref{tbec_ph}(b). Based on the form of the interactions in the Hamiltonian
of the system, the energies of system is degenerate for all the possible
combinations of $\rho^1$ and 
$\rho^2$ in the allowed ranges. For example, with $U_{12}= 0.4U$
and for $\mu/U=0.2$, $J/U=0.01$ we observe $0.33\lesssim \rho^1\lesssim 0.7$.
In the figure, the half filling lobe $\rho^k = 0.5$ and $\rho = 1$ at $J/U=0$
lies in the domain $0\leqslant\mu/U \leqslant 0.4$. In general, in the miscible
domain, the half filling lobe $\rho =1$ at $J/U=0$ lies in the domain
$0\leqslant\mu/U \leqslant U_{12}/U$. The other Mott lobes with higher $\rho$ 
occur at the higher values $\mu/U$. In general, the Mott lobes have 
$\rho=n$ with $n\in\mathbb{N}$ and $\rho^k = n/2$. Thus, for Mott lobes with 
odd $n$ the average occupancy of each species is half integer.

With increasing $U_{12}$, the Mott lobes with odd-integer occupancies grows in 
size, but the size of the lobes with even-integer occupancies remains the same
until $U_{12}=U_{kk}$ but shifts to higher $\mu/U$. This can be understood from
Eq.~(\ref{order_par_ana}). The trend is discernible from the phase diagrams in 
Fig.\ref{tbec_ph}(b)-\ref{tbec_ph}(c). This, in the case of weakly interacting 
binary condensates, is equivalent to a march towards phase separation 
\cite{ho_96,ao_98,gautam_11,roy_15_2,bandyopadhyay_17}. 
For $U_{12}>U_{kk}$, the criterion for phase separation, the 
size of the Mott lobe $\rho =2$ is different. But, once the phase separation
criterion is met, there is no change in the phase diagram with 
further increase in $U_{12}$. As an example the phase diagram for $U_{12}=1.2U$
is shown in  Fig.\ref{tbec_ph}(d). The lobes in this phase diagram are 
the same as in Fig.\ref{tbec_ph}(a). The only difference is the occupancy 
is $\rho=n$ with $n\in\mathbb{N}$ and $\rho^k=n/2$. As a result, the 
density pattern of the lowest Mott lobe ($\rho=1$) has one atom at each 
lattice site chosen randomly from the two species. The important point is that 
the Mott lobes have the same sizes for $U_{12}=0$ and $U_{12}\geqslant U_{kk}$.
But, the occupancy and hence the density patterns are different. 

To verify our results we do a comparison with quantum Monte
Carlo results reported in earlier works \cite{parny_11,kato_14}. For this, we
check the order of the MI-SF quantum phase transition of the $\rho=2$ Mott 
lobe. As a measure we compute the energy per particle for fixed $\mu$ and
find that the transition is first order close to the tip of the Mott lobe.
And, it is second order away from the tip. This is consistent with the quantum
Monte Carlo results \cite{parny_11,kato_14}. To assess the impact of the 
quantum fluctuations on the nature of the phase transitions, we employ 
cluster-Gutzwiller Mean Field (CGMF) theory. This is a multisite 
generalization of the SGMF theory, and captures the quantum correlations 
accurately within each cluster. We refer to 
Refs.~\cite{luhmann_13,bai_18,pal_19} for the details. 
In the present case, we repeat the SGMF computations by using $2\times2$ 
clusters, which is sufficient to probe the effects of quantum fluctuations. 
With the CGMF method, apart from the enhancement of the Mott lobe, we observe 
shrinking in the domain of the first-order phase transition. In particular, 
for $U_{12} = 0.8U$ and $\rho = 2$ Mott lobe, the first-order MI-SF phase 
transition is observed for $1.1\lesssim \mu/U\lesssim 1.3$ with the CGMF 
calculation. While with SGMF, it is $0.99 \lesssim \mu/U\lesssim 1.44$.
Similar trends were reported in the comparison of the mean-field theory and
quantum Monte Carlo results in Ref.~\cite{kato_14}.

%%%%%%%%%%%%%%%%%%%%%%%%%%%%%%%%%%%%%%%%%%%%%%%%%%%%%%%%%%%%%%%%%%%%%%%%%%%%
%%%% subsection: Finite temperature
%%%%%%%%%%%%%%%%%%%%%%%%%%%%%%%%%%%%%%%%%%%%%%%%%%%%%%%%%%%%%%%%%%%%%%%%%%%%
\subsection{Phase diagram at finite temperatures}
\label{fint_phd_tbhm}
\begin{figure}[t]
  \begin{center}
  \includegraphics[width=8.0cm]{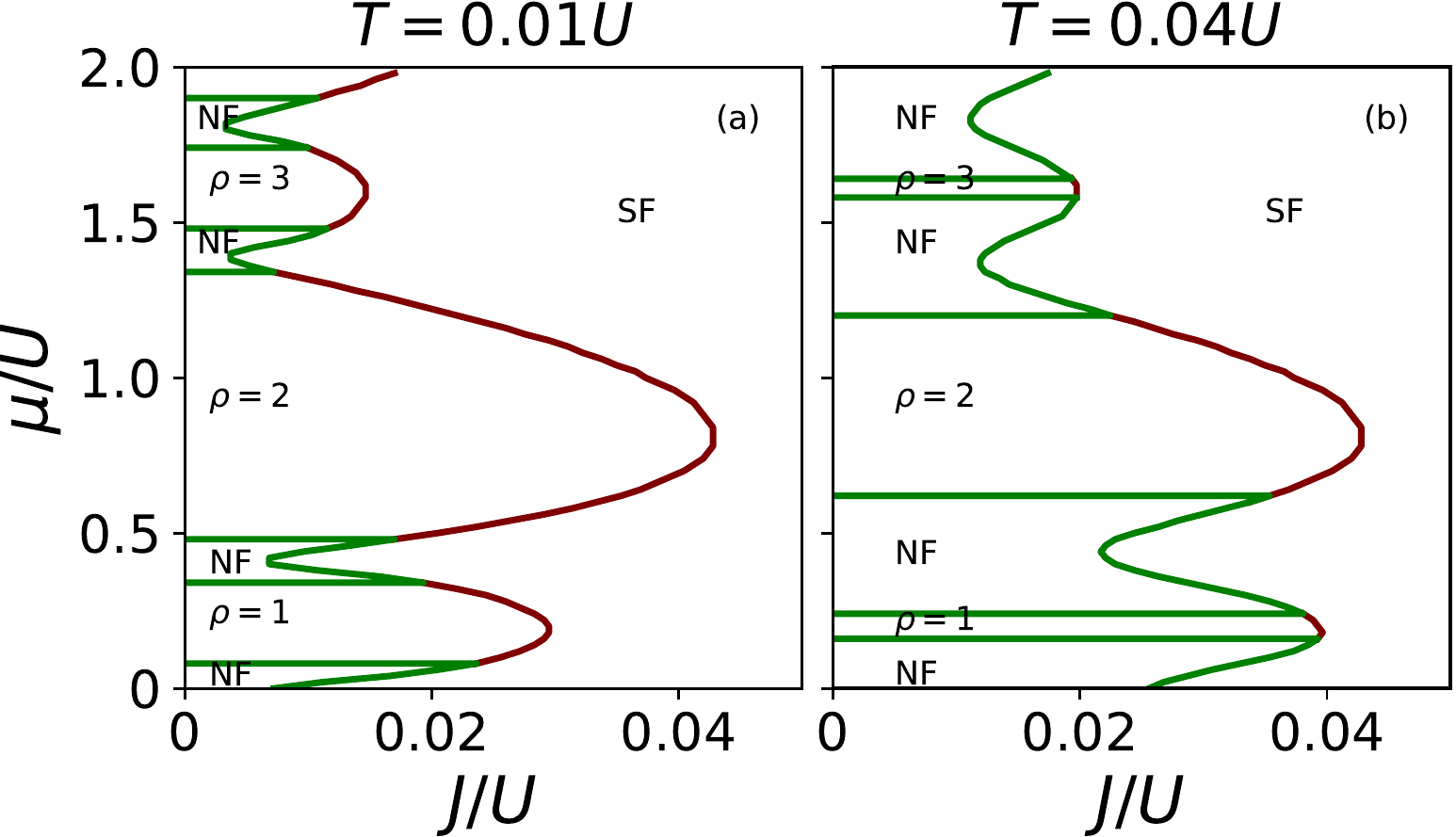}
	  \caption{Phase diagram of TBHM at different
           temperatures for $U_{12} = 0.4 U$.
           Maroon colored line represents the phase boundaries between the
           incompressible lobes and the superfluid phase, while the green line
	   forms the boundary of region of the normal fluid (NF) phase.}
  \label{phd_fint}
  \end{center}
\end{figure}
 The results we have discussed are at zero temperature, a theoretical 
simplification. This simplification helps to explore the basic qualitative 
features of the quantum phases in the system. In these results the thermal 
fluctuations are absent. Experiments are, however, at finite temperatures
and effects of thermal fluctuations have to be incorporated. The competition 
between the quantum and thermal fluctuations modify the zero-temperature phase 
diagram. At finite temperatures, the observables have to be calculated with 
the thermal averaging, and this requires calculation of the partition function. 
In the mean-field theory we have used, the single-site partition function
\begin{equation}
  Z_{p,q} = \sum_{l} e^{-\beta E_{p,q}^l},
\end{equation}
where $\beta = 1/k_{B}T$, $T$ is the temperature of the system, and 
$E_{p,q}^l$ is the $l$th eigenenergy of the single site Hamiltonian at the
lattice site $(p,q)$. As the parameters in the Hamiltonian in 
Eq.(\ref{ham_ss_tbec}) are scaled with the onsite interaction $U$, the 
temperature of the system is in the units of $U/k_B$. And, for simplicity, 
we set $k_B = 1$. The thermal average of the superfluid order parameter for the 
$k$th species at the $(p,q)$ lattice site is
\begin{equation}
     \langle \phi_{p,q}^{k}\rangle = \frac{1}{Z_{p,q}}\sum_{l}
                                               \prescript{l}{p,q}{\bra{\psi}}
                              \hat{b}_{p,q}^{k}
                              e^{-\beta E^l_{p,q}} \ket{\psi}^l_{p,q},
\end{equation}
where $\langle\ldots\rangle$ represents the thermal averaging and 
$\ket{\psi}^l_{p,q}$ is the $l$th eigenstate of the single site Hamiltonian. 
Similarly, the occupancy or the density at finite $T$ is defined as
\begin{equation}
   \langle \langle \hat{n}_{p,q}^{k} \rangle \rangle = \frac{1}{Z_{p,q}}
                                             \sum_{l}
                                             \prescript{l}{p,q}{\bra{\psi}}
                           \hat{n}_{p,q}^{k} e^{-\beta E^{l}_{p,q}}
                           \ket{\psi}^l_{p,q}.
\end{equation}
For a detail implementation of the finite temperature Gutzwiller method, we
refer to Refs. \cite{suthar_20, pal_19}.  At finite temperatures, there is an 
additional phase, normal fluid phase, in the phase diagram. It emerges due to 
the thermal fluctuations \cite{mahmud_11, parny_12}. This phase has superfluid 
order parameter $\phi = 0$, and the local density is real. To distinguish the 
normal fluid phase from the incompressible $\rho = n$ lobes, we compute the 
local compressibility $\kappa$, which is proportional to the local number 
variance. The $\kappa$ is zero in the incompressible phase, while it is finite 
for the normal fluid phase. As an example, in the Fig. \ref{phd_fint}, we show 
the phase diagrams of the TBHM at $T=0.01U$ and $0.04U$ with the same 
interaction parameters as in Fig.\ref{tbec_ph}(b). That is with the 
interspecies interaction $U_{12}=0.4U$. In the phase diagrams, the 
thermal-fluctuations-induced melting of the incompressible lobes into normal
fluid phase is visible. At finite temperature, the normal fluid phase occupies
the regions with $\mu$ below and above the tip of the lobes. The domain of this 
phase is enhanced as the temperature is increased and this is evident from the
Fig.~\ref{phd_fint}. This results in the shrinking of the incompressible lobes.
Upon increasing the temperature further, the incompressible phases disappear
above a critical temperature. For the parameters considered, $T\approx 0.061U$
is the critical temperature at which the incompressible lobes completely melt.

%%%%%%%%%%%%%%%%%%%%%%%%%%%%%%%%%%%%%%%%%%%%%%%%%%%%%%%%%%%%%%%%%%%%%%%%%%%%%%%
%%%% Section: Phase diagram with long range interactions
%%%%%%%%%%%%%%%%%%%%%%%%%%%%%%%%%%%%%%%%%%%%%%%%%%%%%%%%%%%%%%%%%%%%%%%%%%%%%%%

\section{Phase diagram with long-range interactions} \label{sec_phdiag_ext}
\subsection{$V_{12}=0$}

\begin{figure}[t]
  \begin{center}
  \includegraphics[width=8.0cm]{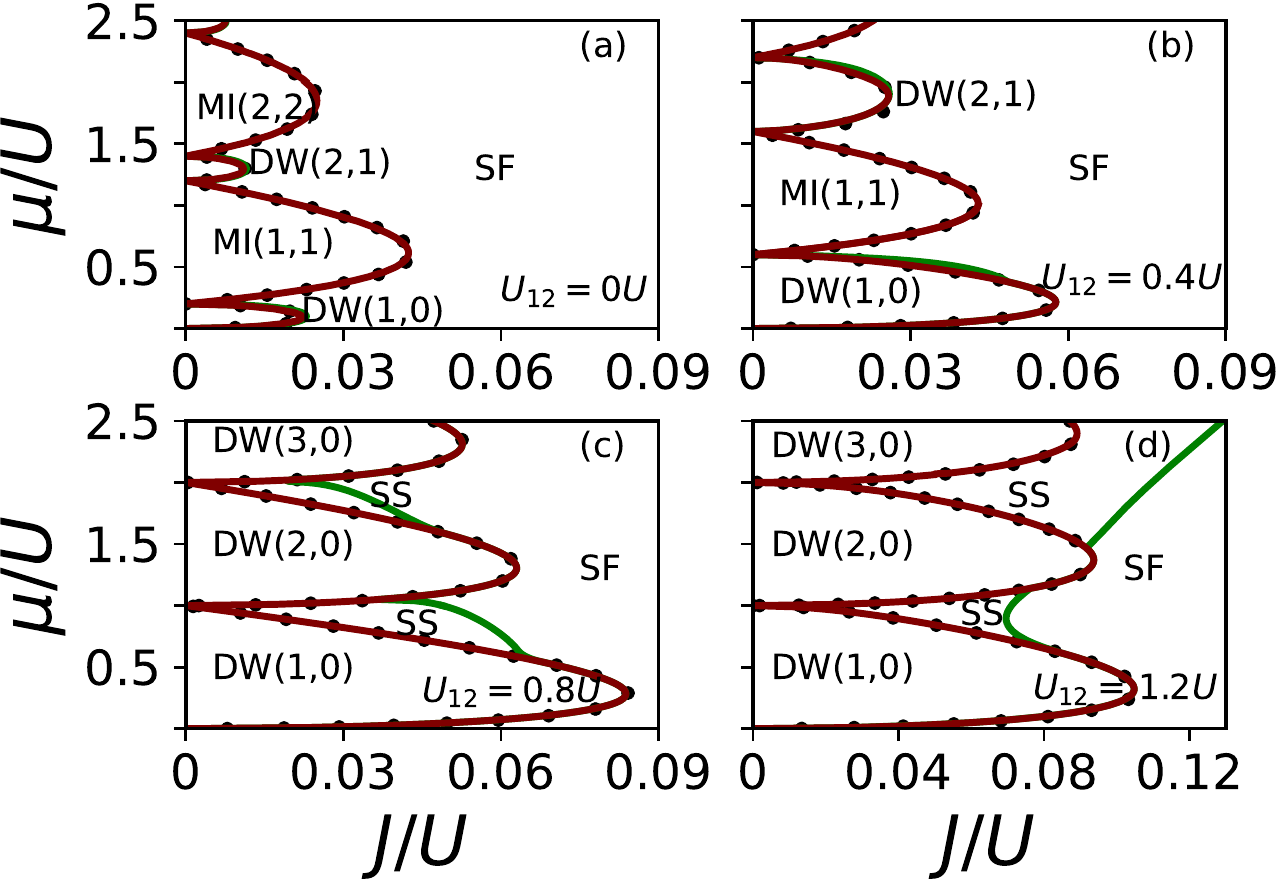}
  \caption{Phase diagram of eTBHM at different interspecies interaction 
	   strength $U_{12}$ and for interspecies NN interaction 
	   $V_{12} = V_{21} = 0$, $V_1 = V_2 = 0.05U$.
           Maroon-colored line forms the boundary of the region comprised of
	   incompressible phases (MI, DW).
	   The filled black dots mark the phase boundaries
	   obtained analytically by perturbative analysis of
           the mean field Hamiltonian. Around the density wave phase,
	   a supersolid phase exists and the boundary
           between supersolid and superfluid phases is represented by green 
	   lines. The supersolid region around the density wave region
	   enlarges with increasing $U_{12}$.
           In DW(n,0) phase both species have DW(n,0) pattern and in MI(1,1)
           phase both species have uniform unit occupancy.}
  \label{ebhm_v12_0}
  \end{center}
\end{figure}

The ground state of the eTBHM Hamiltonian in Eq.~(\ref{bhm_ext_ss}),
like in the previous case, is by obtained using the Gutzwiller ansatz. 
The long-range interactions in the eTBHM introduce two more phases, density
wave and supersolid, in the phase diagram. To analyze and highlight the effect 
of long-range intra- and interspecies interactions, we first consider the 
case of $V_{12} = 0$. And we set the intraspecies NN interaction strength
$V_k = 0.05U$. Then, we vary the interspecies onsite interaction strength 
$U_{12}$, which can be achieved in experiments through the Feshbach resonance. 
The choice of low value of $V_k$ is based on the parameters realized in 
dipolar Bose-Einstein Condensate experiments~\cite{baier_16}.
In these experiments, $V/\hbar$ is in the range $\approx 10-100$ Hz, whereas
$U/\hbar$ has typical values in kHz. In addition, this choice of parameters
has the unique possibility to study the MI-DW quantum phase transition by
changing $U_{12}$ and keeping $V_k$ fixed. This is to be contrasted with the
extended Bose-Hubbard model, where the NN interaction strength
$V\ge0.25U$~\cite{iskin_11,suthar_20} marks the critical point for such
quantum phase transitions. Like in the case of the Bose-Hubbard model, we
consider symmetric hopping $J_x^k = J_y^k = J$, identical chemical potential 
$\tilde{\mu}^k_{p,q} = \mu$, and $U_{kk} = U$. The phase diagram for 
$U_{12} =0$ is shown in Fig.~\ref{ebhm_v12_0} (a). It is identical to the 
phase diagram of the single species extended Bose-Hubbard model 
\cite{iskin_11,suthar_20} and consists of the DW(1,0), MI(1,1), DW(2,1), 
MI(2,2), supersolid (green line) and superfluid phases. In the figure, the 
supersolid phase occurs as a thin envelope around the density wave lobes. 
On increasing $V_k$ but keeping the other parameters fixed, the size of the 
density wave lobes and the accompanying envelope of the supersolid phase are 
enhanced. However, the Mott lobes disappear from the phase diagram. This is due
to the higher energy cost of having commensurate occupancy due to the 
intraspecies NN interaction. The same effect is reported in the single species
extended Bose-Hubbard model~\cite{iskin_11,suthar_20}.

%%%%%%%%%%%%%%%%%%%%%%%%%%%%%%%%%%%%%%%%%%%%%%%%%%%%%%%%%%%%%%%%%%%%%%%%%%%%%%%
%%%%%%%%%%%%%%%%%%%%%%%%%%%%%%%%%%%%%%%%%%%%%%%%%%%%%%%%%%%%%%%%%%%%%%%%%%%%%%%

The density wave phases with $U_{12} =0$ are fourfold degenerate. Two of 
the states have $\Delta \rho^1=\Delta\rho^2$ and the other two have  
$\Delta \rho^1=-\Delta\rho^2$. For both set of states, one of the 
degenerate states is obtained by shifting both of the species by one 
lattice constant either along the $x$ or $y$ direction. For the 
$\Delta \rho^1=\Delta\rho^2$ states, the occupancies of the two species at 
each lattice sites are the same $n^1_{p,q}=n^2_{p,q}$. From this the 
$\Delta \rho^1=-\Delta\rho^2$ states are obtained after translation 
of one of the species by one lattice constant either along the
$x$ or $y$ direction. Thus, in the latter we have $n^{1,A}=n^{2,B}$
and $n^{1,B}=n^{2,A}$. It is to be noted that the $\rho=1$ phase of the TBHM 
has the same average density as the DW(1,0). However, the two have different 
symmetries. The $\rho=1$ phase of the TBHM  has atoms from the two 
species with random occupancies and has no diagonal long-range order.
But, the DW(1,0) has diagonal order arising from the nonzero $\Delta\rho^k$.
As an example, consider the DW(1,0) phase, the two degenerate states 
correspond to $\Delta\rho^1=\Delta\rho^2=1$ and 
$\Delta\rho^1=-\Delta\rho^2=1$. At higher $\mu$, the DW(2,1) intervenes 
the transition from MI(1,1) to MI(2,2) phase.

To study the effect of the interspecies interaction we increase $U_{12}$, 
retaining $V_{12}$ and $V_k$ fixed at $0$ and $0.05U$, respectively. The 
phase diagram corresponding to $U_{12} = 0.4U$ is shown in 
Fig.~\ref{ebhm_v12_0} (b). At finite $U_{12}$ the Mott insulator phase is 
energetically costly due to repulsion between atoms of the two-species 
coexisting on the same lattice site. So it shifts to higher 
$\mu/U$ values with increasing $U_{12}$ which can be understood from 
Eq.~(\ref{exTBHM_MI2}). As seen from the figure, the finite $U_{12}$ 
enhances the DW(1,0) lobe. The finite $U_{12}$ also lifts the degeneracy of 
the density wave states, and the state with  $n^1_{p,q}=n^2_{p,q}$ has 
higher energy. So, the density of the density wave states with finite 
$U_{12}$ has $n^{1,A}=n^{2,B}$ and $n^{1,B}=n^{2,A}$. 

The MI(1,1) lobe remains unchanged in size but is shifted upward in the 
phase diagram. The shift is attributed to the increase in effective chemical
potential arising from the interaction energy associated with finite $U_{12}$. 
A similar trend, enhancement of the DW(1,0) lobe, occurs in the case of
$U_{12}=0$ upon increasing $V_k$. In addition to the Mott insulator phase,
the DW(2,1) and similar density wave phases with nonzero $n^{k,A}$ and
$n^{k,B}$ are also energetically disfavoured. However, the most important
feature is the emergence of prominent supersolid phase envelope around each of
the density wave lobes. Upon increasing $U_{12}$ further, as seen from the 
Figs.~\ref{ebhm_v12_0} (c)-\ref{ebhm_v12_0}(d), the Mott lobes are transformed
into density wave lobes. And, at higher $U_{12}$, only the
DW(n,0) phase, with $n\in\mathbb{N}$, is present in the system. The domain of 
the supersolid phase also increases. Ultimately, the supersolid envelopes 
around each of density wave lobes merge into a single large supersolid domain, 
and this is discernible in these figures.

%%%%%%%%%%%%%%%%%%%%%%%%%%%%%%%%%%%%%%%%%%%%%%%%%%%%%%%%%%%%%%%%%%%%%%%%%%%%
%%%%   Subsection: $V_{12}>0$
%%%%%%%%%%%%%%%%%%%%%%%%%%%%%%%%%%%%%%%%%%%%%%%%%%%%%%%%%%%%%%%%%%%%%%%%%%%%

\subsection{$V_{12}>0$}
\begin{figure}[t]
  \begin{center}
  \includegraphics[width=8.0cm]{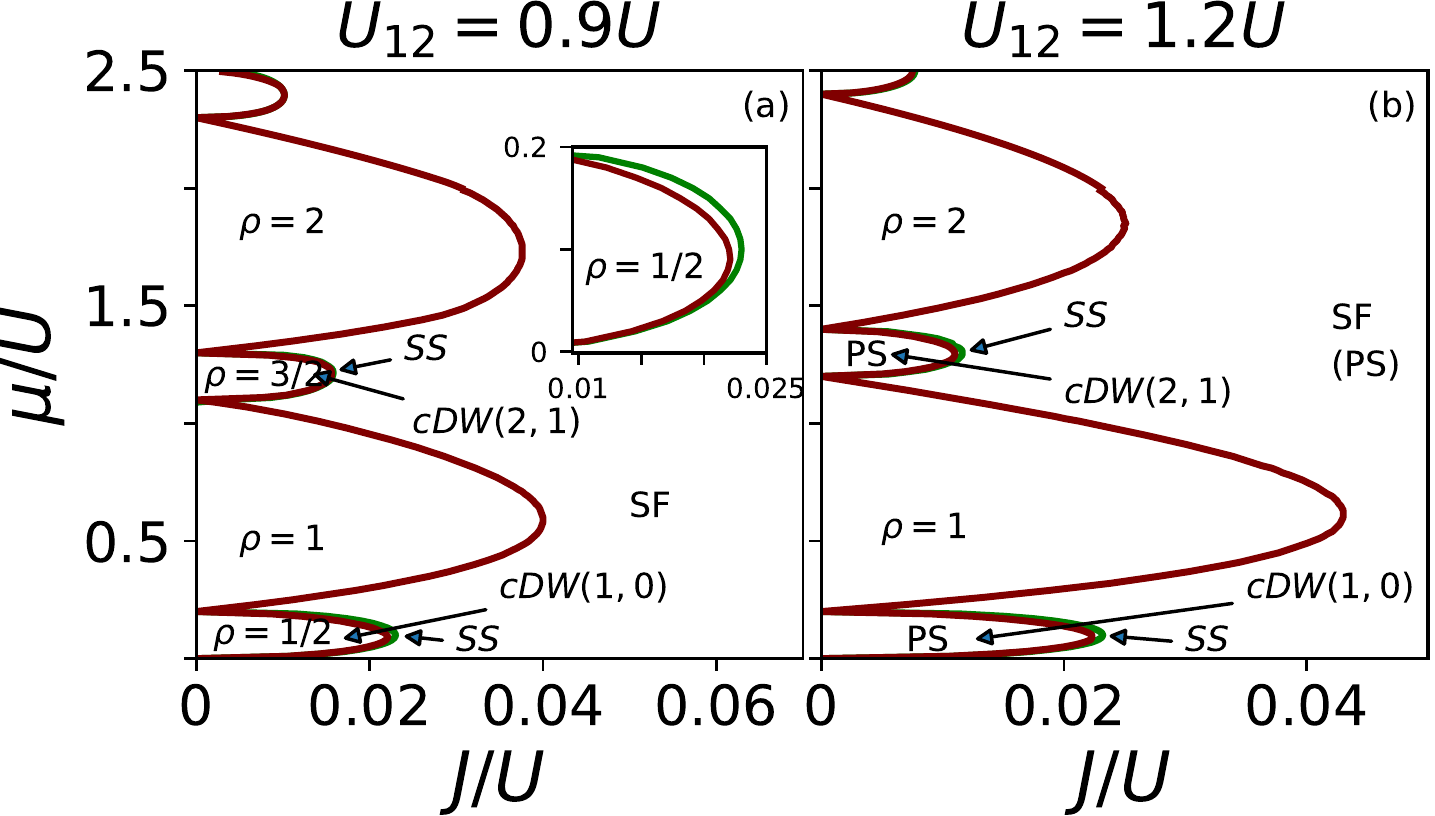}
  \caption{Phase diagram of eTBHM at the different interspecies interaction 
	   strength $U_{12}$ and for interspecies NN interaction 
	   $V_{12} = V_{21} = 0.05$, $V_1 = V_2 = 0.05U$.
           The incompressible (MI, cDW) and compressible phase (SS, SF) regions
           are separated by maroon colored lines. In correlated density wave 
	   phase the two species occupy lattice sites randomly in such a way 
	   such that total density $\rho= \rho^1 +\rho^2$ have density wave 
	   pattern. And, around this phase, the supersolid phase exists and its
	   boundary is marked by green lines. For $U_{12} = 1.2$ the density
	   wave and superfluid phases are phase separated.}
  \label{ebhm_v12_0p05}
  \end{center}
\end{figure}

One of the phenomena unique to the binary condensate is the phase separation. 
This provides important insights to understand novel phenomena in 
nonlinear dynamics, pattern formation, quantum phase transitions in
condensed-matter systems, etc.
\cite{gautam_10_1,gautam_10_2,gautam_11,roy_15_2,
bandyopadhyay_17,gautam_12,gautam_13,roy_14_2,kuopanportti_19,roy_14_1,
suthar_15,roy_15_1,lee_16,suthar_16,suthar_17,pal_17,pal_18}.
Phase separation of binary condensates in the weakly interacting regime, 
as mentioned earlier, is well studied. This, however, is not the case for 
the strongly interacting two-species ultracold atoms in optical lattices. 
As discussed earlier, in the TBHM we observe phase separation in the superfluid
phase, where the density of the two species are spatially separated into two 
domains. The phase-separated Mott insulator phases, on the other hand, have
random filling of the two species and are not separated into two distinct 
domains. The inclusion of the NN interactions modifies its density distribution 
in the phase separated domain. To study this, we solve the 
Eq.~(\ref{bhm_ext_ss})  with finite $V_{12}$ and keep it fixed to a value 
of $0.05U$. We, then, increase the interspecies interaction $U_{12}$ from the 
miscible domain $U_{12}^2 < U_{11}U_{22}$ to the immiscible domain 
$U_{12}^2 > U_{11}U_{22}$. The phase diagrams for selected values of $U_{12}$  
are shown in the Fig.~\ref{ebhm_v12_0p05}.

%%%%%%%%%%%%%%%%%%%%%%%%%%%%%%%%%%%%%%%%%%%%%%%%%%%%%%%%%%%%%%%%%%%%%%%%%%%%
%%%%%%%%%%%%%%%%%%%%%%%%%%%%%%%%%%%%%%%%%%%%%%%%%%%%%%%%%%%%%%%%%%%%%%%%%%%%

%%%%%%%%%%%%%%%%%%%%%%%%%%%%%%%%%%%%%%%%%%%%%%%%%%%%%%%%%%%%%%%%%%%%%%%%%%%%
%%%% Subsubsection: Miscible phase
%%%%%%%%%%%%%%%%%%%%%%%%%%%%%%%%%%%%%%%%%%%%%%%%%%%%%%%%%%%%%%%%%%%%%%%%%%%%

\subsubsection{Miscible phase}
In the miscible domain, $U_{12}^2 < U_{11}U_{22}$, the phase diagram has
lobes of incompressible quantum phases having $\rho=n$ with $n\in \mathbb{N}$.
These lobes are similar to those in the TBHM. In the present case, however, the 
$\rho=n$ lobes are intervened by lobes of density wave quantum phases with 
half-integer total average occupancies $\rho = (2m+1)/2$ with 
$m\in \{0,\mathbb{N}\}$. The total occupancy  $n_{p,q} = n^1_{p,q} + n^2_{p,q}$
of these phases have diagonal long-range order. This is essentially induced by 
the nonzero interspecies NN interaction, $V_{12}>0$. The particle densities 
$n^k_{p,q}$, however, possess no diagonal long-range order. For this reason 
we refer to these as correlated density wave (cDW) phases. This is to 
distinguish between the density wave phases with $V_{12}=0$, in which case 
$n^k_{p,q}$ have diagonal long-range order. Due to the small value of the NN 
interaction strength, the correlated density wave lobes are surrounded by a 
thin envelope of the supersolid phase. As an example, the phase diagram for 
$U_{12}=0.9U$ is shown in Fig.~\ref{ebhm_v12_0p05}(a). 
In the figure, the cDW(1,0) has the lowest average occupancy $\rho=1/2$. 
One of the possible density distributions of this phase is $n^{k,A}=0$. 
And, at the other sublattice the occupancy is 
$n^{2,B}_{p,q} = 1 - n^{1,B}_{p,q}$. The values of $n^{1,B}_{p,q}$ are either 
0 or 1, distributed randomly. And, the random distribution implies that there is
no diagonal long-range order. In other words, the lattice occupancies of the 
individual species are not structured but the total lattice occupancy is a 
structured quantum phase. Around the correlated density wave phase, as $J/U$ is
increased for fixed $\mu/U$, the quantum fluctuations drive a second-order 
quantum phase transition from correlated density wave to the supersolid phase. 
For the supersolid phase around the cDW(1,0) phase, the occupancies of the two 
sublattices are identical, and lie in the range
$0\lesssim n^{1,A}_{p,q}=n^{2,A}_{p,q}\lesssim 0.25$ and 
$0.25\lesssim n^{1,B}_{p,q}=n^{2,B}_{p,q}\lesssim 0.50$. Hence, both the 
species have the same diagonal long-range orders. Here, the occupancies  
are defined over a finite range due to its finite 
compressibility. The superfluid order parameters, although different in value, 
follow similar trends $\phi^{1,A}_{p,q}=\phi^{2,A}_{p,q}$, 
$\phi^{1,B}_{p,q}=\phi^{2,B}_{p,q}$ and 
$\phi^{k,B}_{p,q}\neq\phi^{k,A}_{p,q}$. In short, the fluctuations drive the 
cDW(1,0) phase with random integer $n^{k,B}_{p,q}$ to identical occupancies. 
And, $n^{k,A}_{p,q}$ also acquire nonzero values. Upon increasing $J/U$ 
further, the quantum fluctuations drive another phase transition from the
supersolid to the superfluid phase. In this transition, the diagonal long-range 
order is destroyed and translational invariance of the system is restored. 

The insulating phase with average occupancy $\rho=1$, has uniform total
lattice occupancy $n_{p,q}=n^1_{p,q} + n^2_{p,q}=1$. And, the occupancies of 
the two species satisfy the condition $n^1_{p,q}=1-n^2_{p,q}$ with 
$n^2_{p,q}\in \{ 0,1\}$, where the values between the two possibilities
are chosen at random. Thus, this phase is like the conventional Mott insulator 
phase with integer commensurate integer occupancies, but in terms of the 
total occupancy $n_{p,q}$. Similar to the correlated density wave phase, we
refer to this phase as the correlated Mott insulator phase. This implies that
increasing the chemical potential or adding more particles to the system, at
a fixed but low $J/U$, the system starting from cDW(1,0) passes through
supersolid, superfluid and then to the $\rho=1$ phase. At still higher $\mu$,
the cDW(2,1) phase appears. The total occupancies of the two sublattices in
this quantum phase are $n^A_{p,q} = n^{1,A}_{p,q} + n^{2,A}_{p,q} =2$ and 
$n^B_{p,q} = n^{1,B}_{p,q} + n^{2,B}_{p,q} =1$. This implies that 
both species have the same occupancies in the $A$ sublattice
$n^{1,A}_{p,q} = n^{2,A}_{p,q} = 1$. And, it is equivalent to the 
DW(2,0) phase in the eTBHM with $V_{12}=0$. From this phase we obtain the
cDW(2,1) phase by randomly adding one atom of either species at the 
$B$ sublattice sites. That is, $n^{1,B}_{p,q}=1-n^{2,B}_{p,q}$ with 
$n^{2,B}_{p,q}\in \{ 0,1\}$, where the values between the two possibilities
are chosen at random. So, effectively, the cDW(2,1) is a superposition of 
DW(2,0) with cDW(1,0). At higher $\mu$ the other lobes with 
increasing $\rho$ appear. And these have similar occupancies and 
order parameter structure as the lobes with lower $\rho$. It is to be
highlighted that the phase diagrams are different, qualitatively and
quantitatively, from the two-species Bose-Hubbard model where
only one of the species is dipolar \cite{wilson_16}.

The effect of quantum fluctuations are underestimated in the single-site
mean-field theory. And this could lead to the appearance of quantum
phases which are rendered unstable by quantum fluctuations. The supersolid 
quantum phase, with diagonal long-range order, is one such phase. So, to 
check the robustness of the thin supersolid phase around the correlated density
wave phase, we use the CGMF theory, with which we study the ground state
quantum phases by tiling the system with $2\times 2$ clusters. With this
method, we observe an enhancement of the incompressible lobes. And the extent
of the supersolid phase around the cDW(1,0) phase is similar in size. We also
observe the enhancement of cDW(1,0) lobe along the $\mu/U$ axis. That is, the
cDW (1,0) lobe closes at $\mu = 0.3U$ with CGMF, as compared with $\mu = 0.2U$
calculated by using SGMF. Thus, the supersolid quantum phase around the 
correlated density wave phases appears to be robust against quantum 
fluctuations. A concrete observation could be made with larger clusters and by 
doing a detailed study with cluster finite-size analysis. We shall take this up
in our future works.

%%%%%%%%%%%%%%%%%%%%%%%%%%%%%%%%%%%%%%%%%%%%%%%%%%%%%%%%%%%%%%%%%%%%%%%%%%%%%%%

%%%%%%%%%%%%%%%%%%%%%%%%%%%%%%%%%%%%%%%%%%%%%%%%%%%%%%%%%%%%%%%%%%%%%%%%%%%%
%%%% Subsubsection: Immiscible phase
%%%%%%%%%%%%%%%%%%%%%%%%%%%%%%%%%%%%%%%%%%%%%%%%%%%%%%%%%%%%%%%%%%%%%%%%%%%%
\subsubsection{Immiscible phase}
 The criterion for phase separation of the two species in the binary 
condensates or weakly interacting domain is $U_{12}^2>U_{11}U_{22}$ 
\cite{ho_96,ao_98}. And, as 
discussed earlier, at phase separation the atoms of different species
do not occupy the same lattice site. This is the energetically 
favorable configuration. However, the local nature of the interparticle
interaction preserves the inversion symmetry and the species do not separate
into two spatial domains. In the binary condensates or weakly interacting 
domain, the contact interaction is sufficient to break the inversion 
symmetry and leads to the formation of two spatial 
domains \cite{papp_08,tojo_10,mccarron_11,
wacker_15,wang_16} at phase separation. The introduction of the long-range 
interspecies interaction ($V_{12}>0$) in the eTBHM introduces the 
possibility to lower the energy of the density configurations which breaks 
inversion symmetry. Thus, there is phase ordering of the two species. 

\begin{figure}[t]
  \begin{center}
  \includegraphics[width=8.0cm]{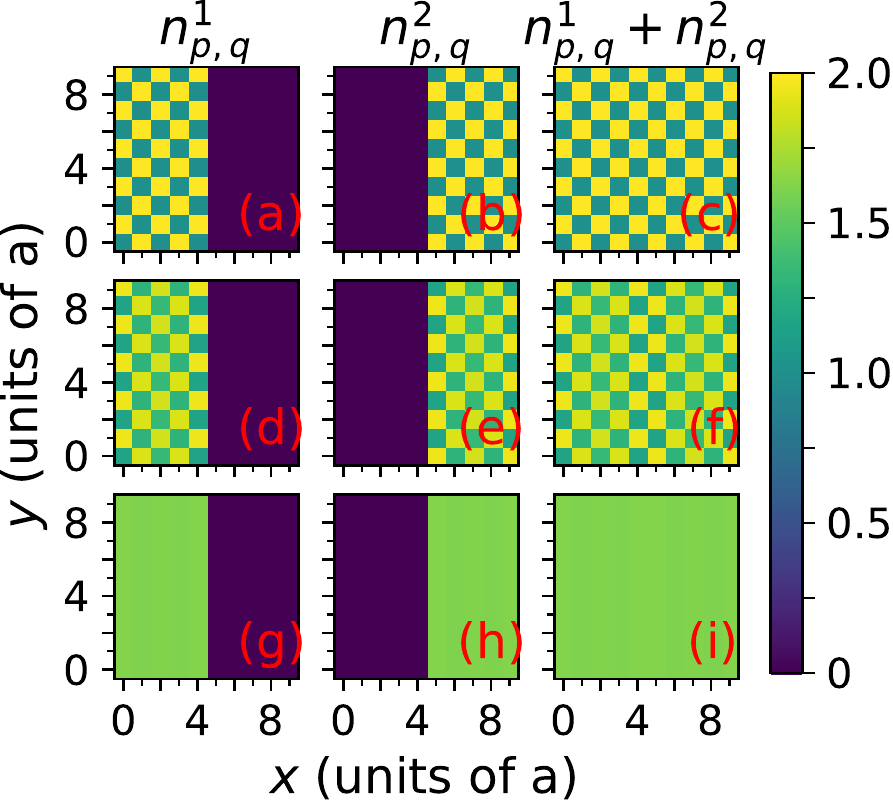}
  \caption{Phase separation with a side-by-side pattern of species occupancies,
           obtained with periodic boundary conditions along both the $x$ and
	   $y$ axes. The density distribution of the species over lattice sites
	   is shown in panels (a)-(c) for the cDW(2,1) phase, in panels (d)-(f)
	   for the supersolid phase, and in panels (g)-(i) for the superfluid
	   phase.}
  \label{dw_ss_sf_occ}
  \end{center}
\end{figure}

 In the present case, for the parameters considered ($U_{11}=U_{22}=U$), the 
phase separation criterion is equivalent to $U_{12}>U$. This choice of 
parameters, as a representative case, capture the key qualitative and 
quantitative features of the eTBHM. More importantly, the long-range nature of 
$V_{12}$ introduces phase ordering at phase separation. As an example, the 
phase diagram for $U_{12}=1.2U$ is shown in  Fig.~\ref{ebhm_v12_0p05}(b). 
The structure of the insulating or incompressible and  compressible phases are 
similar to the case of $U_{12}<U$, as shown in Fig.~\ref{ebhm_v12_0p05}(a). 
But, there is one key difference, the correlated density wave, supersolid and 
superfluid phases in Fig.~\ref{ebhm_v12_0p05}(b) are phase separated. This is 
the combined effect of the onsite and long-range interspecies interactions. 
And, this is indicated in the phase diagram with the annotation PS 
(phase separated). But the insulating phases with $\rho=1$ and $\rho=2$ 
are not phase separated. In the $\rho=1$ phase, like in the case of $U_{12}<U$,
each lattice site is singly occupied by an atom from the two species chosen 
randomly. If the phase separation is along one of the axes, say the $x$-axis, 
the DW($n_A, n_B$) phase has occupancies
\begin{equation}
   n^k_{p,q}=
      \begin{cases}
	      \Theta\left [(-1)^k (p-(K-1)/2)\right ]n_A 
	      		&  \text{for $(p,q) \in$ A} \\
         \Theta\left [(-1)^k (p-(K-1)/2)\right ]n_B 
	      		& \text{for $(p,q) \in$  B},
      \end{cases}
   \label{dw_ps_stripe1}
\end{equation}
where $k$, as defined earlier, is the species index,
$K$ is the size of the system along the $x$-axis, and $n_A$ and $n_B$ are 
integers with $n_A\neq n_B$. The ground state is doubly degenerate because the
above density configuration has the same energy when the 
species are interchanged. The occupancies of other phase-separated phases 
can also be defined in a similar way. However, in these two phases 
$n_A$ and $n_B$ are real. Furthermore, in the supersolid phase $n_A\neq n_B$ but
in the superfluid phase $n_A= n_B$. The superfluid order parameters for these 
phases are also defined in the same form. The presence of the Heaviside step 
functions in Eq. (\ref{dw_ps_stripe1}) indicates
inversion symmetry is broken. The Hamiltonian is, however, invariant under 
the inversion symmetry. Thus, the phase mixed to separation transition
breaks the inversion symmetry spontaneously. And the observed ground state
is one of the degenerate configurations.  

\begin{figure}[t]
  \begin{center}
  \includegraphics[width=8.0cm]{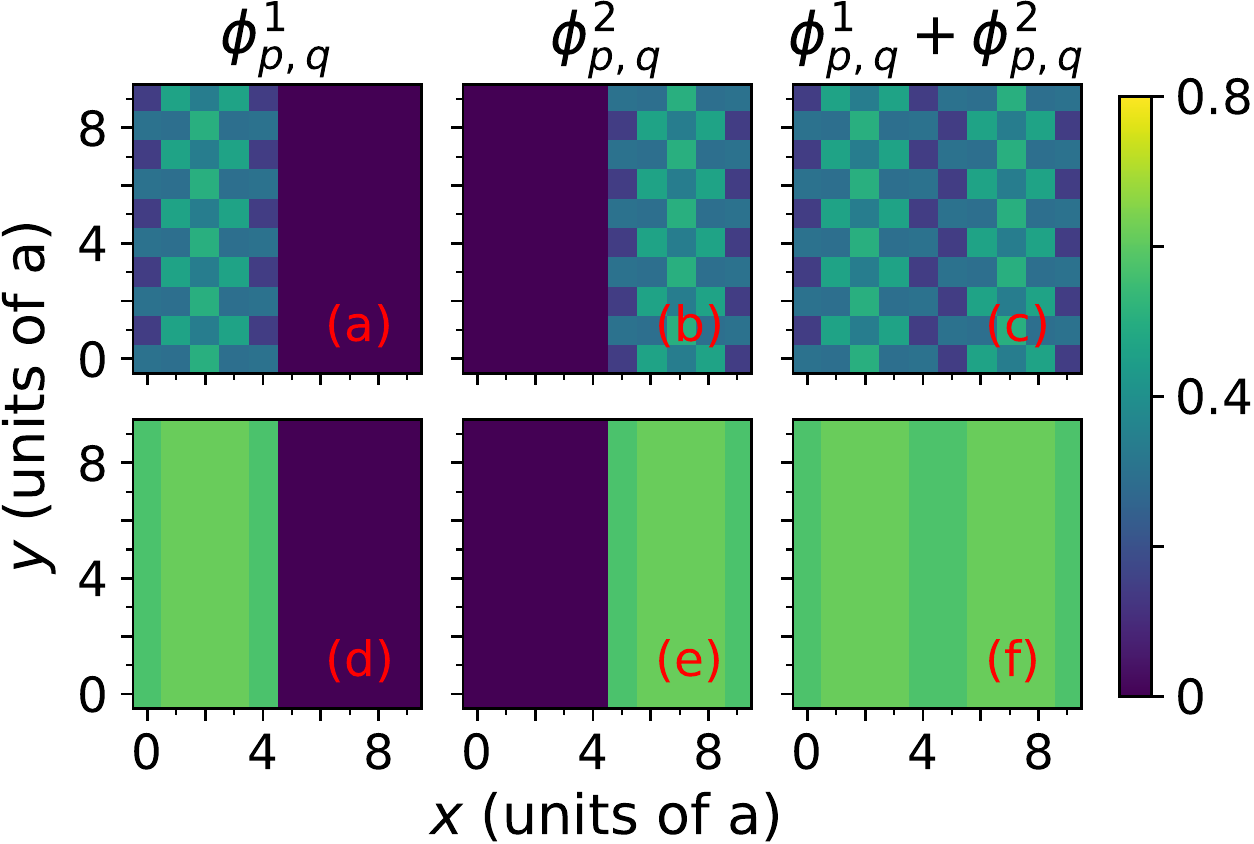}
  \caption{Phase separation with side-by-side pattern of species superfluid 
           order parameter, obtained with periodic boundary conditions along 
	   both $x$ and $y$ axes. The superfluid order parameter at the 
	   lattice sites is shown in panels (a)-(c) for supersolid phase and in 
	   panels (d)-(f) for the superfluid phase.}
  \label{ss_sf_phi}
  \end{center}
\end{figure}

 As an example, the phase diagram in the immiscible domain $U_{12} = 1.2 U$
is shown in Fig.~\ref{ebhm_v12_0p05} (b). In the phase diagram, the global
features of the phase domains are qualitatively similar to the phase diagram in
the miscible parameter domain $U_{12} = 0.9 U$ shown in 
Fig.~\ref{ebhm_v12_0p05} (a). There is, however, an important difference. 
All the phases in the figure are phase separated and this is indicated in
the phase diagram with label PS. In the superfluid phase, phase separation
occurs across the whole parameter domain. The density profiles of the cDW(2,1)
phase, and the supersolid and superfluid phases around it are shown in 
Fig.(\ref{dw_ss_sf_occ}). In the figure, consider the lattice sites with 
odd (even) values of $(p+q)$ as the $A$ ($B$) sublattice. And, for better 
representation of the density orders of the structured phases, we consider a 
system size of $10\times 10$. Then, from the density pattern in 
Figs.(\ref{dw_ss_sf_occ})(a) and \ref{dw_ss_sf_occ}(b), both the species have
occupancies $n_A=2$ and $n_B=1$. And, because it is phase separated, from 
Eq. (\ref{dw_ps_stripe1}), 
\begin{equation}
   n^k_{p,q}=
     \begin{cases}
         \Theta\left [(-1)^k(p-\frac{9}{2}) \right ]2 &  
                 \text{for odd $(p+q)$} \\
         \Theta\left [(-1)^k(p-\frac{9}{2}) \right ]1 &  
                 \text{for even $(p+q)$}.
      \end{cases}
   \label{dw21_sp12}
\end{equation}
The density pattern shown in the figures 
Figs.(\ref{dw_ss_sf_occ})(a)-\ref{dw_ss_sf_occ}(c) 
correspond to the parameters $\mu/U=1.35 $ and $J/U=0.010$. The above 
occupancies of the species imply that each of the species are confined within
a subsystem of a $5\times 10$ lattice. The other species, because we apply
periodic boundary conditions along both directions, effectively 
provides a confining potential. This is better visualized when the system
is mapped to a torus. Then, phase separation along one of the axes, divides
the torus into two equal halves. Here each half is occupied by one of the 
species. For such a configuration, there are two interspecies boundaries 
which segregate the two species. Thus, with a $10\times 10$ system size,
the total length of the boundary is $20a$, where $a$ (as defined earlier) is
the lattice constant. From the figure it is evident that other 
configuration is the phase-separated state, existing along the diagonal. This,
however, is energetically not favourable as it has larger interface energy due
to longer boundary $10(2+\sqrt{2})a$.

For the same value of chemical potential $\mu/U=1.35$, on increasing the
hopping amplitude to $J/U=0.011$ we are in the supersolid phase domain. It is 
also phase separated and the lattice site occupancies have a similar form
as Eq.(\ref{dw21_sp12}). The occupancies are real, have checkerboard
order, and are shown in Fig.(\ref{dw_ss_sf_occ})(d)-\ref{dw_ss_sf_occ}(f). 
Another important point is, as seen from the figures, that boundary effects are
present in the superfluid order parameter. The reason is that the effective 
potential which segregates the two species is like a soft boundary
condition. And, this is due to the long-range interspecies interaction. 
The supersolid phase is a superfluid phase with diagonal long-range order, and 
hence has a nonzero superfluid order parameter $\phi^k_{p,q}$. The 
superfluid order parameters of the two species are shown 
in Fig.\ref{ss_sf_phi}(a)-\ref{ss_sf_phi}(c). The 
boundary effects are more prominent in these figures and at the boundaries, 
the deviations from the checkerboard order of $\phi^k_{p,q}$ are visible 
without ambiguity. It is to be mentioned here that the domain of the supersolid 
phase, for the parameters considered, is rather small. Despite this, supersolid 
quantum phase with phase segregation is a novel one and it deserves
detailed investigations. Upon increasing $J/U$ further, we reach the superfluid
phase, which is also phase separated. As an example, the occupancies 
and superfluid order parameters for $\mu/U=1.35$ and $J/U=0.015$ are shown in 
Figs.(\ref{dw_ss_sf_occ})(g)-\ref{dw_ss_sf_occ}(i)
and in Figs.\ref{ss_sf_phi}(d)-\ref{ss_sf_phi}(f), respectively. 
In the superfluid phase, there is phase separation, but the occupancies and 
superfluid order parameter are uniform within the domains of each species. 
Thus, the average occupancies and lattice site occupancies are the same
$n^1_{p,q}=\rho^1$ (for $p<5$) and $n^2_{p,q}=\rho^2$ (for $p \geqslant 5$).
As we consider identical parameters for both the species
$\rho=\rho^1=\rho^2$, where $\rho\in \mathbb{R}$ and
$1\leqslant \rho\leqslant 2$. The values and range are also discernible
from the figures. The key point from these case studies is that, for 
nonzero interspecies long-range interactions and $U_{12}^2>U_{11}U_{22}$, 
the eTBHM has quantum phases which are phase separated.

%%%%%%%%%%%%%%%%%%%%%%%%%%%%%%%%%%%%%%%%%%%%%%%%%%%%%%%%%%%%%%%%%%%%%%%%%%%%
%%%% Subsubsection: Finite tilt angle and finite temperature
%%%%%%%%%%%%%%%%%%%%%%%%%%%%%%%%%%%%%%%%%%%%%%%%%%%%%%%%%%%%%%%%%%%%%%%%%%%%

\subsubsection{Finite tilt angle and finite temperature}
 \label {fint_tilt_ang}
 The results of the eTBHM discussed so far are the quantum phases of the model
described by the Hamiltonian in Eq.\ref{bhm_ext}. As mentioned earlier, this 
corresponds to the case of dipoles aligned perpendicular to the lattice 
plane. In this section we provide a brief discussion on the general case, 
where the tilt angle $\theta$ is nonzero. For this, we consider the 
dipole-dipole interaction
\begin{equation*}
  \frac{C_{\rm dd}}{2}\sum_{ij}\hat{n}_i\hat{n}_j
  \frac{(1-3\text{cos}^2\alpha_{ij})}{|\mathbf{r}_i - \mathbf{r}_j|^3},
\end{equation*}
where the angle $\alpha_{ij}$ is the angle between the dipole polarization 
axis and the separation vector $\mathbf{r}_i - \mathbf{r}_j$ between the
lattice sites $i$ and $j$.  The coupling constant $C_{\rm dd}$
represents the strength of the dipole interaction. Without loss of generality,
the dipoles are assumed to be polarized in the $y-z$ plane, and then, 
$\theta = \frac{\pi}{2} - \alpha_{ij}$. The detailed physical
description of such a system is given in Ref. \cite{bandyopadhyay_19}.  
Even though the dipole-dipole interaction is a long-range interaction, we 
restrict it to the NN sites. This simplified limit is sufficient to examine 
the effects arising from the anisotropy of the interaction.
The strength of the dipole-dipole interaction can be varied from 
$C_{\text{dd}}$ to $-2 C_{\text{dd}}$, by changing $\theta$ from
0 to $\frac{\pi}{2}$. Thus the {\em effective} dipole-dipole interaction 
strength decreases as $\theta$ increases. In the repulsive domain, the 
decrease in the effective interaction strength shrinks the density wave and 
Mott lobes. We have verified the decrease of the density wave lobes by 
considering the tilt angle $\theta = \frac{\pi}{12}$. In the miscible phase 
($U_{12} = 0.9U$), the phase diagram at $\theta = \frac{\pi}{12}$ is 
qualitatively similar, but there are quantitative differences in terms of the 
phase boundaries of the incompressible phases. As stated earlier, we observe 
that the correlated density wave lobes shrink along the $J/U$ axis and close at
a smaller $\mu/U$ value. The thin envelope of the supersolid phase also show 
the same trends as the correlated density wave lobes. That is, the supersolid
phase also shrinks along the $J/U$ axis, and closes at a lesser $\mu/U$ value.
The incompressible $\rho = n$ lobes are shifted downward along the $\mu$ axis. 

  Earlier, we had discussed the ground-state phases with thermal fluctuations
associated with finite temperatures. We, similarly, have studied the effects 
of the thermal fluctuations on the ground-state phases of eTBHM, in particular,
for the parameters domain where the system is in the miscible domain. We 
observe that the regions of the incompressible lobes are reduced, and the 
melted region is occupied by the normal fluid phase. Like in the case of TBHM, 
in Sec. \ref{fint_phd_tbhm}, an increase in temperature shrinks the 
incompressible lobes. And, above a critical temperature, the lobes disappear.

%%%%%%%%%%%%%%%%%%%%%%%%%%%%%%%%%%%%%%%%%%%%%%%%%%%%%%%%%%%%%%%%%%%%%%%%%%%%
%%%% Subsubsection: Stability analysis of the phases                       %
%%%%%%%%%%%%%%%%%%%%%%%%%%%%%%%%%%%%%%%%%%%%%%%%%%%%%%%%%%%%%%%%%%%%%%%%%%%%

\subsubsection{Linear Stability analysis}
 \label {stab_anly}

 The dynamics of fluid mixtures exhibit different types of instabilities. The
binary condensates are no exception. In particular, the Rayleigh-Taylor 
instability \cite{sasaki_09,gautam_10_1} and Kelvin-Helmholtz instability 
\cite{takeuchi_10,lundh_12} have been studied in detail. So, it is pertinent to
examine the stability of the spatially phase-separated ground-state 
configuration of the eTBHM. The collective excitations are the relevant 
properties of the system which carry signatures of instabilities. To calculate
the collective excitations we add fluctuations $\delta c^{(p,q)}_{n_1,n_2}(t)$
to the ground-state coefficients in the dynamical Gutzwiller mean-field
equation \cite{jakub_05,kovrizhin_05}. The coefficients of the Fock states in 
Eq. \ref{gw_2s} is then modified to 
\begin{equation}
   c^{(p,q)}_{n_1,n_2}(t) = \bar{c}^{(p,q)}_{n_1,n_2} 
                            + \delta c^{(p,q)}_{n_1,n_2}(t),
\end{equation}
where $\bar{c}^{(p,q)}_{n_1,n_2}$ are the coefficients at equilibrium or 
the ground-state solution of the Gutzwiller mean-field theory. To obtain the
collective excitations, we use the Bogoliubov approximation and define
\begin{equation}
  \delta c^{(p,q)}_{n_1,n_2}(t) = u_{n_1,n_2}^{(p,q)}e^{-i\omega t} 
                + v_{n_1,n_2}^{*(p,q)}e^{i\omega t},
  \label{delta_c}
\end{equation} 
where $\omega$ is the energy of the collective mode, and 
$(u_{n_1,n_2}, v_{n_1,n_2})$ is the
amplitude of the collective modes \cite{krutitsky_10, krutitsky_11, saito_12}. 
Using this in the dynamical Gutzwiller equation, and retaining terms linear
in $u$ and $v$, we obtain the Bogoliubov-de Gennes equation. The details of
the derivation and equations for the eTBHM are given in
Appendix \ref{appendix_c}. We, then, diagonalize the Bogoliubov-de Gennes 
matrix and obtain the eigenspectrum of the system.

In the eigenspectrum of the system, the appearance of collective modes with 
complex energies is a signature of dynamical instability. With complex 
energy, the imaginary part leads to an exponential growth of the fluctuations
and this is evident from Eq.(\ref{delta_c}). And, thus, the system is unstable
to perturbations. To determine the stability of the phases in the phase diagram 
presented in Fig.~\ref{ebhm_v12_0p05} (b), we have performed the stability 
analysis for the phase-separated, side-by-side ordered cDW (2,1) and  
superfluid phases. In both of these phases, we get a real-valued
excitation spectrum. This indicates that these phase-separated states are 
dynamically stable. We have also verified the stability of other phases in the 
phase diagram.

%%%%%%%%%%%%%%%%%%%%%%%%%%%%%%%%%%%%%%%%%%%%%%%%%%%%%%%%%%%%%%%%%%%%%%%%%%%%%%%
%%%% Section: Conclusions
%%%%%%%%%%%%%%%%%%%%%%%%%%%%%%%%%%%%%%%%%%%%%%%%%%%%%%%%%%%%%%%%%%%%%%%%%%%%%%%

\section{Conclusions} \label{sec_conclude}
  In conclusion, we obtain the phase diagram of the two-species 
Bose-Hubbard model and its extended version, the eTBHM with long-range 
interactions in two-dimensional optical lattices. Our findings are pertinent 
and timely in view of the recent experimental realization of the Er-Dy binary 
dipolar Bose-Einstein condensate mixture \cite{trautmann_18}. The phase diagram
of the TBHM  has the unique feature of additional Mott lobes with average 
occupancies which are half integer. These lobes emerge due to the presence of 
the second species. And, the domain of these lobes are enhanced with the 
increase of the interspecies interaction strength. In the case of eTBHM, we 
obtain insulating phases with the nonoverlapping density distributions even 
with $U_{12}^2<U_{11}U_{22}$, where the atoms of the two species are
distributed across the system randomly. The nonoverlapping densities
are like phase separation but, in this work, we use phase separation to mean
the configuration where the densities of the two species are segregated 
into two nonoverlapping domains. One key finding of our study is that the 
DW-MI quantum phase transitions may occur by varying $U_{12}$ while keeping 
$V_{k}$ fixed. This is in contrast with the single species extended Bose-Hubbard
model, where the NN interaction strength is required to be large to 
observe such quantum phase transitions. With finite interspecies NN 
interactions, we obtain the phase diagram in the miscible and immiscible 
regimes. Our result is that the correlated density wave, supersolid, and 
superfluid phases in the eTBHM in the immiscible domain $U_{12}^2>U_{11}U_{22}$
are phase separated. And, they have side by side order. 
These phase-separated phases are dynamically stable.

%%%%%%%%%%%%%%%%%%%%%%%%%%%%%%%%%%%%%%%%%%%%%%%%%%%%%%%%%%%%%%%%%%%%%%%%%%%%%%%
%%%% Section: Acknowledgements
%%%%%%%%%%%%%%%%%%%%%%%%%%%%%%%%%%%%%%%%%%%%%%%%%%%%%%%%%%%%%%%%%%%%%%%%%%%%%%%

\section{Acknowledgements} \label{sec_acknowledge}
 The results presented in the paper are based on computations using
Vikram-100, the 100TFLOP HPC Cluster at the Physical Research Laboratory,
Ahmedabad, India. K.S. acknowledges the support of the National Science
Centre, Poland via project 2016/21/B/ST2/01086.

%%%%%%%%%%%%%%%%%%%%%%%%%%%%%%%%%%%%%%%%%%%%%%%%%%%%%%%%%%%%%%%%%%%%%%%%%%%%%
%%%% Appendix: A
%%%%%%%%%%%%%%%%%%%%%%%%%%%%%%%%%%%%%%%%%%%%%%%%%%%%%%%%%%%%%%%%%%%%%%%%%%%%%

\appendix
\section{Perturbation Analysis of the Two-Species Bose-Hubbard Model}\label{appendix_a}

The unperturbed ground state at the lattice site $(p,q)$ has the form 
$|\psi\rangle_{p, q}^{(0)} =|n^1,n^2\rangle_{p, q}$. The energy of this 
unperturbed ground state is 
\begin{eqnarray}
E^{(0)}_{n^{1}_{p,q},n^{2}_{p,q}} 
	&=& \frac{U}{2} \left[n^{1}_{p,q}(n^{1}_{p,q} -1)+n^{2}_{p,q}
	(n^{2}_{p,q} -1)\right]\nonumber\\
	&+& U_{12} n^{1}_{p,q} n^{2}_{p,q} -\mu^{1}_{p,q}n^{1}_{p,q} 
	-\mu^{2}_{p,q}n^{2}_{p,q},
 \label{unperturb_energy}
\end{eqnarray}
where we have chosen $U_{11}=U_{22}=U$. Then, to the first order of the 
superfluid order parameter $\phi^{k}_{p,q}$ the perturbed ground state can 
be written as
\begin{eqnarray}
\!\!\!\!\!\!\!\!\!\!\!\!|\psi\rangle_{p,q} 
	        &=& |n^1,n^2\rangle_{p,q} \nonumber\\
		&+& \!\!\!\!\!
		\sum_{\substack{m^1,m^2\\ \neq n^1,n^2}} 
		\frac{_{p,q}\langle m^1,m^2| \hat{h}_{p,q,1}^{{\rm TBH}}
                |n^1,n^2 \rangle_{p,q}}
		{E^0_{n^1_{p,q},n^2_{p,q}} -E^0_{m^1_{p,q},m^2_{p,q}}} 
		|m^1,m^2\rangle_{p,q},%\nonumber \\
 \label{perturb_gs}			
\end{eqnarray} 
where, considering uniform hopping strengths for both the species 
($J_x^1 =J_x^2 =J_y^1 =J_y^2 =J$) and superfluid order parameters as 
real numbers
\begin{eqnarray}
\!\!\!\!\!\!\!\!\!\!\!\!\hat{h}_{p,q,1}^{{\rm TBH}} =
	        -J \left[ \bar{\phi}_{p, q}^1 \left(
		\hat{b}_{p, q}^{\dagger 1} +\hat{b}_{p, q}^1 \right)
		+ \bar{\phi}_{p, q}^2 \left(\hat{b}_{p, q}^{\dagger 2} 
		+ \hat{b}_{p, q}^2 \right)\right],
\label{hop_perturb}
\end{eqnarray}
with $\bar{\phi}^{k}_{p,q} = \left(\phi^{k}_{p+1,q}+\phi^{k}_{p-1,q}
+\phi^{k}_{p,q+1}+\phi^{k}_{p,q-1}\right)$. Then, using 
Eqs.~(\ref{unperturb_energy})--(\ref{hop_perturb}) the ground state
can be calculated as
\begin{eqnarray}
|\psi\rangle_{p,q} =&& |n^1,n^2\rangle_{p,q} \nonumber \\
		&&+J \bar{\phi}^1_{p,q} 
		\left\{\frac{\sqrt{n^1_{p,q} +1}}
		{n^{1}_{p,q}U -\mu^{1}_{p,q} +U_{12}n^{2}_{p,q}}
		|n^{1} +1,n^{2}\rangle_{p,q} \right.\nonumber \\
		&&\left.- \frac{\sqrt{n^1_{p,q}}}{(n^1_{p,q}-1)U 
		-\mu^{1}_{p,q} +U_{12}n^2_{p,q}}
		|n^1 -1,n^2 \rangle_{p,q}\right\} \nonumber\\
		&&+J \bar{\phi}^2_{p,q} \left\{\frac{\sqrt{n^2_{p,q} +1}}
		{n^2_{p,q}U -\mu^{2}_{p,q} +U_{12}n^1_{p,q}}
		|n^1,n^2 +1\rangle_{p,q} \right.\nonumber \\
		&&\left.-\frac{\sqrt{n^2_{p,q}}}{(n^2_{p,q}-1)U 
		-\mu^{2}_{p,q} +U_{12}n^1_{p,q}}
		|n^1,n^2 -1\rangle_{p,q}\right\}. \nonumber \\
 \label{perturb_psi}
\end{eqnarray}

From this state, we calculate the superfluid order parameter 
$\phi^{k}_{p,q} = \, _{p,q}\langle\psi|\hat{b}^{k}_{p,q}|\psi\rangle_{p,q}$,
and the expression is given in Eq.~(\ref{order_par_ana}).

%%%%%%%%%%%%%%%%%%%%%%%%%%%%%%%%%%%%%%%%%%%%%%%%%%%%%%%%%%%%%%%%%%%%%%%%%%%%%
%%%% Appendix: B
%%%%%%%%%%%%%%%%%%%%%%%%%%%%%%%%%%%%%%%%%%%%%%%%%%%%%%%%%%%%%%%%%%%%%%%%%%%%%

\section{Perturbation analysis of the Extended Two-Species Bose-Hubbard Model}\label{appendix_b}
The unperturbed ground state at the lattice site $(p,q)\in A$ sublattice has 
the form $|\psi\rangle_{A}^{(0)} =|n^{1,A},\,n^{2,A}\rangle$ with energy
\begin{eqnarray}
E^{(0)}_{n^{1,A},n^{2,A}} 
	&=& \frac{U}{2} \left[n^{1,A}(n^{1,A} -1)+n^{2,A}(n^{2,A} -1)\right]
				\nonumber\\
	&+& U_{12}\, n^{1,A} n^{2,A} - \mu (n^{1,A} +n^{2,A})
				\nonumber\\
	&+& 4V_1\, (n^{1,A} n^{1,B} + n^{2,A} n^{2,B})
\label{unperturb_energy_etbhm}
\end{eqnarray}
where we have chosen $\tilde{\mu}^{1} =\tilde{\mu}^{2} =\mu$, 
$U_{11} =U_{22} =U $ and $V_1 =V_2$. Then, to first order in the 
superfluid order parameter $\phi^{k}_{p,q}$, the ground state is
\begin{eqnarray}
 \!\!\!\!\!|\psi\rangle_{A} \quad&& = \quad|n^{1},\,n^{2}\rangle_{A} \quad +
	                 \nonumber\\
                 && \!\!\!\!\!\!\!\!\sum_{\substack{(m^{1},m^{2})\\ 
			  \neq (n^{1},n^{2})}} 
		\frac{_{A}\langle m^{1},m^{2}|\hat{h}_{p,q,1}^{{\rm TBH}}
		|n^{1},n^{2} \rangle_{A}}
		{E^0_{n^{1,A},n^{2,A}} -E^0_{m^{1,A},m^{2,A}}} 
		|m^{1},\,m^{2}\rangle_{A} \,,
\label{perturb_gs_etbhm}			
\end{eqnarray} 

where, considering $J_x^1 =J_x^2 =J_y^1 =J_y^2 =J$ and the superfluid order 
parameters as real numbers
\begin{eqnarray}
\hat{h}_{p,q,1}^{{\rm TBH}}= -4J \left[ \phi_{B}^1 \left(
			\hat{b}_{A}^{\dagger 1} +\hat{b}_{A}^1 \right)
			+ \phi_{B}^2 \left(\hat{b}_{A}^{\dagger 2} 
			+ \hat{b}_{A}^2 \right) \right]. 
\label{hop_perturb_etbhm}
\end{eqnarray}

Then, using Eqs.~(\ref{unperturb_energy_etbhm})--(\ref{hop_perturb_etbhm}) the
perturbed ground state is
\begin{eqnarray}
\!\!\!\!|\psi\rangle_{A}  &=& |n^{1,A},n^{2,A}\rangle \nonumber \\
			  &+& 4J \phi^1_{B} \left\{\frac{\sqrt{n^{1,A} +1}}
			       {Un^{1,A} -\mu +U_{12}n^{2,A} +4V_1 n^{1,B}}
			       |n^{1,A} +1,\, n^{2,A}\rangle \right. \nonumber\\
			&-&\left. \frac{\sqrt{n^{1,A}}}
			       {U(n^{1,A}-1) -\mu +U_{12}n^{2,A} +4V_1 n^{1,B}}
			       |n^{1,A} -1,\, n^{2,A} \rangle\right\}
			       \nonumber \\
			  &+& 4J \phi^2_{B} \left\{\frac{\sqrt{n^{2,A} +1}}
			       {Un^{2,A} -\mu +U_{12}n^{1,A} +4V_1 n^{2,B}}
			       |n^{1,A},\, n^{2,A} +1\rangle \right. \nonumber\\
			&-&\left. \frac{\sqrt{n^{2,A}}}
			       {U(n^{2,A}-1) -\mu +U_{12}n^{1,A} +4V_1 n^{2,B}}
			       |n^{1,A},\, n^{2,A} -1\rangle\right\}.
			       \nonumber \\
\label{perturb_psi_etbhm}
\end{eqnarray}
Using this, the superfluid order parameter 
$\phi^{1}_{A} = _{A}\langle\psi|\hat{b}^{1}_{A}|\psi\rangle_{A}$ is 
given by
\begin{eqnarray}
\phi^1_{A} &=&  4J \phi^1_{B} \left\{\frac{n^{1,A} +1}
		{Un^{1,A} -\mu +U_{12}n^{2,A} +4V_1 n^{1,B}} \right.
		\nonumber \\
		&-& \left. \frac{n^{1,A}}
		{U(n^{1,A}-1) -\mu +U_{12}n^{2,A} +4V_1 n^{1,B}}
		\right\}. \nonumber \\
\label{order_par_1A_etbhm}
\end{eqnarray}
A similar analysis can be done at the lattice site $(p,q)\in B$ to obtain
the superfluid order parameter
$\phi^{1}_{B} = _{B}\langle\psi|\hat{b}^{1}_{B}|\psi\rangle_{B}$, and we get
\begin{eqnarray}
\phi^1_{B} &=&  4J \phi^1_{A} \left\{\frac{n^{1,B} +1}
		{Un^{1,B} -\mu +U_{12}n^{2,B} +4V_1 n^{1,A}} \right.
		\nonumber \\
		&-& \left. \frac{n^{1,B}}
		{U(n^{1,B}-1) -\mu +U_{12}n^{2,B} +4V_1 n^{1,A}}
		\right\}. \nonumber \\
\label{order_par_1B_etbhm}
\end{eqnarray}

Substituting $\phi^{1}_{B}$ from Eq.(\ref{order_par_1B_etbhm}) into 
Eq.(\ref{order_par_1A_etbhm}) and then, taking the limit
$\phi^1_{A} \rightarrow 0^{+}$ gives Eq.(\ref{order_par_ana_etbhm}), which 
defines the DW-compressible phase boundary.

%%%%%%%%%%%%%%%%%%%%%%%%%%%%%%%%%%%%%%%%%%%%%%%%%%%%%%%%%%%%%%%%%%%%%%%%%%%%%
%%%% Appendix: C
%%%%%%%%%%%%%%%%%%%%%%%%%%%%%%%%%%%%%%%%%%%%%%%%%%%%%%%%%%%%%%%%%%%%%%%%%%%%%

\section{Bogoliubov-de Gennes equations for eTBHM}\label{appendix_c}
%\allowdisplaybreaks
The Bogoliubov-de Gennes equation, obtained after retaining the linear
terms in the fluctuations and using Bogoliubov approximation, for the eTBHM is
\begin{widetext}
\begin{eqnarray}
  \omega \, u_{n_1,n_2}^{(p,q)}
    &=& \sum_{(p',q'),m_1,m_2} \left(
         A^{(p,q)\;n_1,n_2}_{(p',q')\;m_1,m_2}\; u_{m_1,m_2}^{(p',q')} \quad+
         B^{(p,q)\;n_1,n_2}_{(p',q')\;m_1,m_2}\; v_{m_1,m_2}^{(p',q')}
        \right), \nonumber \\
  -\omega \, v_{n_1,n_2}^{(p,q)}
    &=& \sum_{(p',q'),m_1,m_2} \left(
        B^{*(p,q)\;n_1,n_2}_{(p',q')\;m_1,m_2}\; u_{m_1,m_2}^{(p',q')} \quad+
        A^{*(p,q)\;n_1,n_2}_{(p',q')\;m_1,m_2}\; v_{m_1,m_2}^{(p',q')}
       \right). \nonumber
\end{eqnarray}

It is a set of two coupled equations in terms of the mode amplitudes $u$ and
$v$. The matrix elements in the above equations are

\begin{eqnarray}
%\hspace*{-2cm}
  A^{(p,q),n_1,n_2}_{(p',q'),m_1,m_2} =
        &&\Bigg( \bigg[
            \sum_k \left(\frac{U_{kk}}{2}n_k(n_k-1) -\mu^k n_k 
            +V_k n_k \mathbf{N}^k_{(p,q)} \right) 
            + U_{12} n_1 n_2
            +V_{12} (n_1 \mathbf{N}^2_{(p,q)} + n_2 \mathbf{N}^1_{(p,q)})
            -\omega_0^{(p,q)} \bigg]
            \delta_{n_1,m_1}\delta_{n_2,m_2} \nonumber \\
        &&\quad -J^1 \left[ \sqrt{n_1+1}\,\Phi_{(p,q)}^{1*} 
                 \delta_{m_1,n_1+1}  
                + \sqrt{n_1}\,  \Phi_{(p,q)}^{1} 
                 \delta_{m_1,n_1-1} \right]\delta_{m_2,n_2} \nonumber\\
        &&\quad -J^2 \left[ \sqrt{n_2+1}\,\Phi_{(p,q)}^{2*}
                 \delta_{m_2,n_2+1} 
                + \sqrt{n_2}\,  \Phi_{(p,q)}^{2} 
                 \delta_{m_2,n_2-1} \right]\delta_{m_1,n_1}
          \Bigg)\; \delta_{p',p}\, \delta_{q',q} \nonumber \\
       +&&\bigg(
                -J^1 \sqrt{(n_1+1)(m_1+1)}\, \bar{c}^{*(p',q')}_{m_1+1,m_2} \,
                \bar{c}^{(p,q)}_{n_1+1,n_2} \,
                -J^1 \sqrt{n_1m_1}\, \bar{c}^{*(p',q')}_{m_1-1,m_2}\, 
                \bar{c}^{(p,q)}_{n_1-1,n_2} \nonumber \\ 
        &&\quad -J^2 \sqrt{(n_2+1)(m_2+1)}\, \bar{c}^{*(p',q')}_{m_1,m_2+1} \,
                \bar{c}^{(p,q)}_{n_1,n_2+1} \,
                -J^2 \sqrt{n_2m_2}\, \bar{c}^{*(p',q')}_{m_1,m_2-1}\, 
                \bar{c}^{(p,q)}_{n_1,n_2-1} \nonumber \\
        &&\quad +\, \big[\,V_1 n_1 m_1 + V_2 n_2 m_2 + V_{12}(n_1 m_2 +n_2 m_1)
                    \, \big]
             \bar{c}^{*(p',q')}_{m_1,m_2}\, \bar{c}^{(p,q)}_{n_1,n_2}
          \bigg)
                 (\delta_{p',p \pm 1}\delta_{q',q} +\,
                  \delta_{p',p}\delta_{q',q \pm 1}), \nonumber \\
		  \nonumber \\
%\end{eqnarray}
%\begin{eqnarray}
  B^{(p,q),n_1,n_2}_{(p',q'),m_1,m_2} =
        &&\bigg(
                -J^1 \sqrt{(n_1+1)(m_1)}\, \bar{c}^{(p',q')}_{m_1-1,m_2} \,
                \bar{c}^{(p,q)}_{n_1+1,n_2} \,
                -J^1 \sqrt{n_1(m_1+1)}\, \bar{c}^{(p',q')}_{m_1+1,m_2}\, 
                \bar{c}^{(p,q)}_{n_1-1,n_2} \nonumber \\ 
        &&\quad -J^2 \sqrt{(n_2+1)(m_2)}\, \bar{c}^{(p',q')}_{m_1,m_2-1} \,
                \bar{c}^{(p,q)}_{n_1,n_2+1} \,
                -J^2 \sqrt{n_2(m_2+1)}\, \bar{c}^{(p',q')}_{m_1,m_2+1}\, 
                \bar{c}^{(p,q)}_{n_1,n_2-1} \nonumber \\
        &&\quad +\, \big[\,V_1 n_1 m_1 + V_2 n_2 m_2 + V_{12}(n_1 m_2 +n_2 m_1)
                    \,\big]
             \bar{c}^{(p',q')}_{m_1,m_2}\, \bar{c}^{(p,q)}_{n_1,n_2}
          \bigg)
                 (\delta_{p',p \pm 1}\delta_{q',q} +\,
                  \delta_{p',p}\delta_{q',q \pm 1}). \nonumber  
\end{eqnarray}
\end{widetext}
Here $k$ represents the species index. $\Phi_{(p,q)}^{k}$ and 
$\mathbf{N}^k_{(p,q)}$ are the mean-field superfluid order parameter and the 
number density summed over NN sites of $(p,q)$ respectively.
And $\omega_{0}^{(p,q)}$ is the ground-state energy 
calculated using the unperturbed coefficients. The equations can be written as
a matrix equation

\begin{equation*}
  \omega
            \begin{pmatrix}
              \mathbf{u} \\
              \mathbf{v}
            \end{pmatrix} =
            \begin{pmatrix}
              \mathbf{A}  & \mathbf{B} \\
              -\mathbf{B^*} & -\mathbf{A^*}
           \end{pmatrix}
           \begin{pmatrix}
              \mathbf{u} \\
              \mathbf{v}
            \end{pmatrix}.
\end{equation*}
The matrix on the right-hand side is the Bogoliubov-de Gennes matrix. 
Diagonalizing, we get the collective modes of the system.

%%%%%%%%%%%%%%%%%%%%%%%%%%%%%%%%%%%%%%%%%%%%%%%%%%%%%%%%%%%%%%%%%%%%%%%%%%%%
%%%%  Bibliography
%%%%%%%%%%%%%%%%%%%%%%%%%%%%%%%%%%%%%%%%%%%%%%%%%%%%%%%%%%%%%%%%%%%%%%%%%%%%

\bibliography{ref}{}

\end{document}